\documentclass{emulateapj}

\newcommand{\MgII}{\hbox{{\rm Mg}{\sc \,ii}}}
\newcommand{\Arelcorr}{\hbox{$0.67 \pm 0.09$}}
\newcommand{\LRGminmass}{\hbox{$3.5\times10^{12}$}}
\newcommand{\LRGmassrangemin}{$\sim2$--$8\times 10^{11}$\,\msun}

\newcommand{\overestimate}{\hbox{25\%~$\pm$~10\%}}

\newcommand{\HI}{\hbox{{\rm H}{\sc i}}}

\newcommand{\Ly}{\hbox{{\rm Ly}$\alpha$}}

\newcommand{\lya}{\hbox{{\rm Ly}$\alpha$}}

\newcommand{\msun}{\hbox{M$_{\odot}$}}
\newcommand{\cmsq}{\hbox{cm$^{-2}$}}
\newcommand{\hMpc}{\hbox{$h^{-1}$~Mpc}}
\newcommand{\dNdz}{\hbox{$\frac{\mathrm d N}{\mathrm d z}$}}
\newcommand{\dNdzDLA}{\hbox{$\frac{\mathrm d N}{\mathrm d z\mathrm d \log M}$}}
\newcommand{\dNdl}{\hbox{$\frac{\mathrm d N}{\mathrm d l}$}}
\newcommand{\Hgamma}{\hbox{${\Gamma({1}/{2})\Gamma([{\gamma-1}]/{2})}/{\Gamma({\gamma}/{2})}$}}

\newcommand{\kms}{\hbox{km~s$^{-1}$}}
\newcommand{\nn}{\nonumber}

\newcommand{\besta}{\hbox{$0.73\pm0.08$}}
\newcommand{\biasratiotrue}{\hbox{$0.771$}}
\newcommand{\massratio}{\hbox{$4.8$}}

\newcommand{\massLBGlog}{\hbox{$11.57$}}
\newcommand{\rnotLBG}{\hbox{$2.85$}}
\newcommand{\massDLAall}{\hbox{$2.12^{+2.96}_{-2.0}\times 10^{11}$}}
\newcommand{\massDLAlog}{\hbox{$11.13^{+0.13}_{-0.13}$}}
\newcommand{\massDLAlogtrue}{\hbox{$11.16$}}
\newcommand{\massDLAtrue}{\hbox{$3.94\times 10^{11}$}}

\slugcomment{Received 2004 November 9; accepted 2005 March 13; publication set to 2005 July 20}

\begin{document}

\title{Measuring the Halo Mass of  $z\sim 3$ Damped Ly-alpha 
Absorbers from the Absorber-Galaxy Cross-correlation}
\author{Nicolas   Bouch\'e}
\affil{Max Planck Institut f\"ur extraterrestrische Physik,  Giessenbachstrasse,  D-85748, Garching, Germany; nbouche@mpe.mpg.de}
\author{Jeffrey P. Gardner} 
\affil{Pittsburgh Supercomputing Center, 4400 5th Avenue   Pittsburgh, PA 15213; gardnerj@psc.edu}
\author{Neal Katz}
\affil{Department of Astronomy, University of Massachusetts-Amherst,  Amherst, MA 01003; nsk@kaka.astro.umass.edu}
\author{David H. Weinberg}
\affil{Department of Astronomy,   Ohio State University, 140 West 18th Avenue, Columbus, OH 43210-1173; dhw@astronomy.ohio-state.edu}
\author{Romeel Dav\'e}
\affil{Astronomy Department, University of Arizona, 933 North Cherry Avenue, Tucson, AZ 85721; rad@as.arizona.edu}
\author{James D. Lowenthal}
\affil{Five College Astronomy Department, Smith College, McConnell Hall, Northampton, MA 01063; james@ast.smith.edu}

\begin{abstract}
We test the reliability of a method to measure the mean halo mass of absorption line systems such as
damped \Ly\ absorbers (DLAs). 
The method  is based on measuring the ratio of the cross-correlation   
between DLAs and  galaxies to the   autocorrelation
of the  galaxies themselves, which  is (in linear theory) the ratio of their bias factor
$\overline b$.
We show that the ratio of the  projected cross- and autocorrelation functions 
($w_{\rm dg}(r_\theta)/w_{\rm gg}(r_\theta)$) is also the ratio of their bias factor
irrespective of the galaxy distribution, provided that one uses the same galaxies for 
$w_{\rm dg}(r_\theta)$ and $w_{\rm gg}(r_\theta)$.
Thus, the method requires only multi-band imaging of DLA fields, and is applicable to all redshifts.
 Here, we focus on $z=3$ DLAs. 
 We demonstrate that  the cross-correlation method  robustly constrains the mean DLA halo mass
using smoothed particle hydrodynamics (SPH) cosmological simulations
that resolve  DLAs and galaxies in halos of mass $M_h\ga 5.2\times 10^{10}$\msun.
 If we use the bias formalism of Mo \& White (2002)
 with the DLA and galaxy mass distributions of these simulations,
 we predict an amplitude   ratio $w_{\rm dg}/w_{\rm gg}$ of \biasratiotrue.
 Direct measurement of these correlation functions from 
the simulations yields $w_{\rm dg}/w_{\rm gg}=\overline b_{\rm DLA}/\overline b_{\rm gal}=\besta$, in excellent agreement
with that prediction. Equivalently, inverting the measured correlation ratio
to infer the (logarithmically) averaged DLA halo mass yields  $\langle\log M_{\rm DLA}(\msun)\rangle\;=\massDLAlog$, in excellent agreement with
the true value in the simulations:  $\langle\log M_{\rm DLA}\rangle=\massDLAlogtrue$
is the probability weighted mean mass of the DLA host halos in the simulations.
 The cross-correlation method thus appear to
yield a robust estimate of the average host halo mass even though the DLAs and the galaxies
occupy a broad mass spectrum of halos, and massive halos contain multiple galaxies with DLAs.
If we consider subsets of the simulated galaxies with high star formation rates
 (representing Lyman break galaxies [LBGs]),
then   both correlations are higher, but their ratio still implies the same DLA host mass, irrespective of 
the galaxy subsamples, i.e., the cross-correlation technique is also reliable. 
The inferred mean   DLA halo mass, $\langle\log M_{\rm DLA}\rangle\;=\massDLAlog$,
is an upper limit since the simulations do not resolve halos less massive than $\sim 10^{10.5}$~\msun.
Thus, our results    imply that the correlation length between DLAs and LBGs is predicted to be, at most,
  $\sim \rnotLBG$~\hMpc\ given that $z=3$ LBGs have a correlation length of $r_0\simeq 4$~\hMpc.
While the small size of current observational samples does not allow strong conclusions,
future measurements of this cross-correlation can definitively distinguish models in which
many DLAs reside in low mass halos from those in which DLAs are massive disks occupying only high mass halos.
\end{abstract}


\section{Introduction}

Damped \Ly\ absorbers (DLAs), which cause the strongest absorption lines found in quasar spectra,
have neutral hydrogen (\HI) column densities greater than $2\times 10^{20}$~cm$^{-2}$.
Their integrated column density distribution implies that  DLAs contain the largest reservoir of \HI\ 
at high redshifts \citep[e.g.,][]{LanzettaK_91a,LanzettaK_95b,EllisonS_01a,PerouxC_03b}.
They contain more  \HI\ than all the absorption line clouds in the \Ly\ forest combined;
and in an $\Omega_M=1$ universe, they contain as much hydrogen as
 the comoving mass density of stars in disk galaxies today.
This led \citet{WolfeA_86a} to put forward the hypothesis that 
DLAs are large, thick gaseous disk galaxies.  This hypothesis has been debated since.
On   one hand,  absorption-line velocity profiles of low-ionization species of DLAs 
 seem to be consistent with those
expected from lines of sight intercepting rotating  thick gaseous disks
\citep{WolfeA_95a,ProchaskaJ_97b,LedouxC_98a}.  
\citet{ProchaskaJ_97b} argue that  
 the most likely rotation velocity is $\sim 225$~\kms, i.e., that DLAs are typically
 as massive ($10^{12}~\msun$) as $L^{*}$($z=0$) galaxies.
On the other hand,
 \citet{McDonaldP_99a} and \citet{HaehneltM_98a,HaehneltM_00a} have shown that a large range of structures 
 and morphologies, rather than a single uniform type of galaxy, can
account for the observed  DLA kinematics.
At least at low redshifts ($z<1$), this is supported by observations 
\citep{LeBrunV_97a,KulkarniV_00a,RaoS_00a}.

Early predictions of  DLA properties  
were made using cosmological simulations \citep{KatzN_96a} 
and semianalytical simulations of galaxy formation  \citep{KauffmannG_96a}.
Then, \citet{GardnerJ_97a} extended the results of \citet{KatzN_96a}
to predict the DLA statistics (e.g., ${\mathrm d N}/{\mathrm d z}$) accounting for the limited resolution of those simulations.
They developed a semianalytical method to correct the numerical predictions for the contribution 
of unresolved low-mass halos and found that
roughly half of these systems reside in halos with circular velocities 
$V_c \geq 100$~\kms, and half in halos with $35\kms \leq V_c \leq100$~\kms.
Interestingly, \citet{GardnerJ_97b} found that 
``a CDM model with $\Omega_0 = 0.4$, $\Omega_\Lambda = 0.6$ gives an acceptable fit to the observed absorption
statistics,'' whereas other models did not match the observations so well.
More recently, \citet{GardnerJ_01a} found 
that there was an anti-correlation between the absorber cross section and the 
projected distance to the nearest galaxy, and that DLAs arise out to 10--15 kpc.
Indeed, they found that the
mean cross section for DLA absorption is     much larger than what one would  estimate   
 based on the collapse of the baryons into a centrifugally supported disk.
To match the observed DLA abundances, they required an extrapolation of the mass function
to small halos down to a cut-off of $V_c = 50$--80~\kms. 
 
Other work, such as that of \citet{MoH_99a}, \citet{HaehneltM_98a,HaehneltM_00a}, \citet{NagamineK_04a}, and \citet{OkoshiK_04a},
 indicates that  DLAs are mostly faint (sub-L$^*_{z=0}$) galaxies in small dark matter halos with $V_c\ll 100$~\kms.
However, the exact fraction of DLAs in such halos is a strong function of resolution, as shown by 
\citet{NagamineK_04a}. \citet{FynboJ_99a} and \citet{SchayeJ_01a} 
used cross section arguments and reached  similar conclusions. For instance,  \citet{SchayeJ_01a} argued that 
 the observed Lyman break galaxy (LBG) number density alone ($n = 0.016~h^3$~ Mpc$^{-3}$ down to  0.1~$L^*$)
    can account for all DLA absorptions at $z\sim3$ if the cross section 
    for DLA absorption is $\pi r^2$ with $r=19~h^{-1}$~kpc, much larger than
    the luminous parts of most LBGs \citep{LowenthalJ_97a}. However,  \citet{SchayeJ_01a} pointed out that 
 the cross section can be much smaller than this, if a fraction of DLA systems arise in outflows or
 if $n$ is much larger (i.e., there are many LBGs or other galaxies not yet detected).
 In the semianalytical models of \citet{MallerA_00a}, DLAs arise from the combined
effects of massive central galaxies and a number of smaller satellites within 100~$h^{-1}$~kpc
in virialized halos.
From all these studies, it appears that the low-mass hypothesis is favored
against the thick gaseous disk model of \citet{WolfeA_86a,WolfeA_95a} and \citet{ProchaskaJ_97a}. 
A strong constraint on the nature of DLA will come from 
a measure  of the typical DLA halo mass.

In order to  constrain the mass of $z\simeq3$ DLAs, 
several groups (\citeauthor{GawiserE_01a}~\citeyear{GawiserE_01a};
 \citeauthor{AdelbergerK_03a}~\citeyear{AdelbergerK_03a}, hereafter ASSP03;
 \citeauthor{BoucheN_03a}~\citeyear{BoucheN_03a},~\citeyear{BoucheN_04c} [hereafter BL04];
\citeauthor{BoucheN_03b}~\citeyear{BoucheN_03b};
J. Cooke et al., 2004, private communication)
are using Lyman break galaxies (LBGs) as large-scale structure tracers
 to measure the DLA-LBG cross-correlation, given that   in  hierarchical galaxy formation models,
different DLA masses  will lead to different  clustering properties with the galaxies around them.
Specifically,   the DLA-galaxy cross-correlation  yields a   
measurement of the dark matter halo mass associated with DLAs relative to that of the galaxies. 
In particular, if the galaxies are less (more)  correlated with the DLAs than with themselves, this will imply
that the halos of DLAs are less (more) massive than those of the galaxies.

The purpose of this paper is to use   cosmological simulations
in order to demonstrate that  cross-correlation techniques  will  uniquely constrain the mean DLA halo mass, 
and to compare the results with   observations. 
The advantage of using cosmological simulations 
is that one can check the reliability of the clustering 
results given that the mean halo mass of any  population is  a known quantity  in the simulations.
 As we  show,  we find that 
   the DLA-galaxy cross-correlation implies a  mean DLA halo mass (logarithmically averaged)
 of $\langle\log M_{\rm DLA}(\msun)\rangle\;\simeq\massDLAlog$ close to the  $\langle\log M_{\rm DLA}(\msun)\rangle=11.16$ expected
 from the DLA halo mass distribution.
The method is generally applicable to any redshifts, but we focus here on $z=3$.

Section~\ref{section:simulation} presents the numerical simulations used in this paper.
Section~\ref{section:clustering} lays the foundations of our clustering analysis.
Our results are presented in section~\ref{section:results} along with a comparison
to current   observational results.
 A discussion of the implications of our results is presented in section~\ref{section:conclusions}.

\section{Simulations}
\label{section:simulation}

We use  the TreeSPH simulations of \citet{KatzN_96b} parallelized by \citet{DaveR_97b},
which   combine smoothed particle hydrodynamics  \citep[SPH;][]{LucyL_77a} with
the tree algorithm for computation of the gravitational force \citep{HernquistL_87a}.
This formulation is completely Lagrangian, i.e., it follows each particle in
space and time. The simulations include dark matter, gas, and stars. 
Dark matter particles are collision-less and  influenced only by gravity, while gas particles
are influenced by pressure gradients and shocks in addition to gravity,
and can   cool radiatively. 
Gas particles  are transformed into collision-less stars
when the following conditions are met: 
the local density reaches a certain threshold ($n_H \geq 0.1$ cm$^{-3}$), and
the particles are  colder than a threshold temperature ($T  \leq 30,000$ K)   and are part
of a Jeans unstable convergent flow   (see \citeauthor{KatzN_96b}~\citeyear{KatzN_96b} for details).
A \citet{MillerG_79a} initial mass function of stars is assumed.
Stars of mass greater than 8~\msun\ become supernovae and inject $10^{51}$ erg~s$^{-1}$
 of pure thermal energy into neighboring gas particles. 
Thus, the star formation rate (SFR) is known for each galaxy.
Photo-ionization by a spatially uniform UV background \citep{HaardtF_96a} is included.

The simulation  was run from redshift $z=49$ to redshift $z=0$
 with the following cosmological parameters:
$\Omega_M=0.4, \;\Omega_\Lambda=0.6, \; h\equiv H_0/(100\;\kms~\hbox{Mpc}^{-1})=0.65, \; \Omega_b=0.02~h^{-2},$
a primordial power spectrum index $\;n=0.93,$ and $\sigma_8=0.8$ for the amplitude of   mass fluctuations.
 In this paper, we use the   $z=3$ output.  
The simulation has $128^3$ dark matter particles and the same number of gas particles in a periodic box
of 22.222~$h^{-1}$~Mpc (comoving) on a side with a gravitational softening  length of 3.5~$h^{-1}$ kpc (Plummer equivalent).
The mass of a dark matter particle is $8.2\times 10^8$~\msun, and
 the mass of a baryonic particle is $1.09\times 10^8$~\msun.
We identify dark matter halos  
  by using a ``friends-of-friends'' algorithm \citep{DavisM_85a} 
with a linking length of 0.173 times the mean interparticle separation.
There are 1770 resolved dark matter halos with a minimum of 64 dark matter particles ($5.2\times10^{10}$~\msun).

\begin{figure*}
\plotone{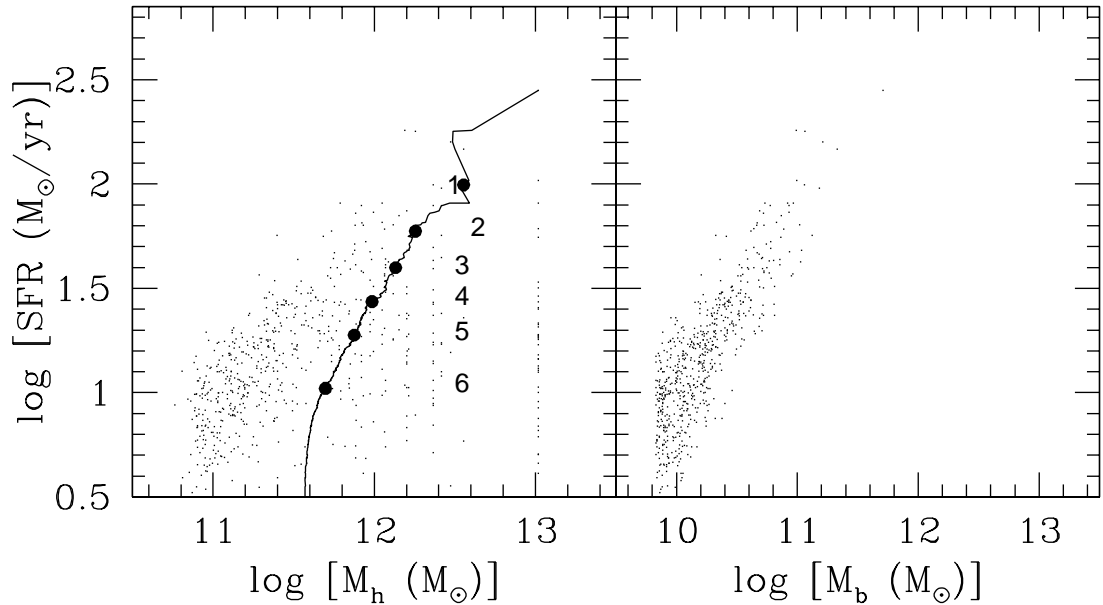}
\caption{{\it Left:} SFR as a function of halo mass (DM~$+$~baryons) $M_h$.
The streak of points at $10^{13}$~\msun\ corresponds to  several resolved galaxies.
 The line shows the running mean (in $\log M$) with a decreasing SFR threshold.
The filled circles  show the SFR threshold vs. the mean mass  ($\langle\log M\rangle$) of
the six subsamples. The six subsamples are the   7, 25, 50,  100, 200, and 400 
  most star-forming galaxies, labeled 1--6.
{\it Right:} SFR as a function of the baryonic mass $M_b$.
\label{cluster:fig:sim:mass}
}
\end{figure*}

We use the group finding algorithm ``spline kernel interpolative denmax''  \citep[SKID;][]{KatzN_96b}
to find   galaxies in the simulations. 
We refer the reader to \citet{KeresD_04a} for a detailed discussion of the SKID algorithm.
There are 651 galaxies resolved 
with  a minimum  of 64 SPH particles (or $6.9\times 10^9$~\msun).
 Figure~\ref{cluster:fig:sim:mass}
shows the SFR  as a function of total halo mass (dark matter $+$ baryons; {\it left})
 and baryonic mass ({\it right})   for the 651 SKID-identified  galaxies. 
The line shows the running mean (in $\log M$) with a decreasing SFR threshold.


The rest-UV spectra and colors of observed LBGs are dominated by the light from massive stars
 \citep{LowenthalJ_97a,PettiniM_01a}.  To simulate various ``flux-simulated''
LBG samples in the simulations, 
 we selected six subsamples of galaxies according to their SFR,
consisting  of the   7, 25, 50,  100, 200, and  400 
  most star-forming galaxies.
 The corresponding SFR
   thresholds and mean masses $\langle\log M\rangle$ for each of the subsamples
   are marked with the filled circles in Figure~\ref{cluster:fig:sim:mass} ({\it left}) labeled 1--6.
Naturally, real LBGs are color-selected, so
this SFR selection can only be an approximation.  \citet{DaveR_99a} discuss
the properties of LBGs in numerical simulations similar to this one.

\begin{figure*}
\plotone{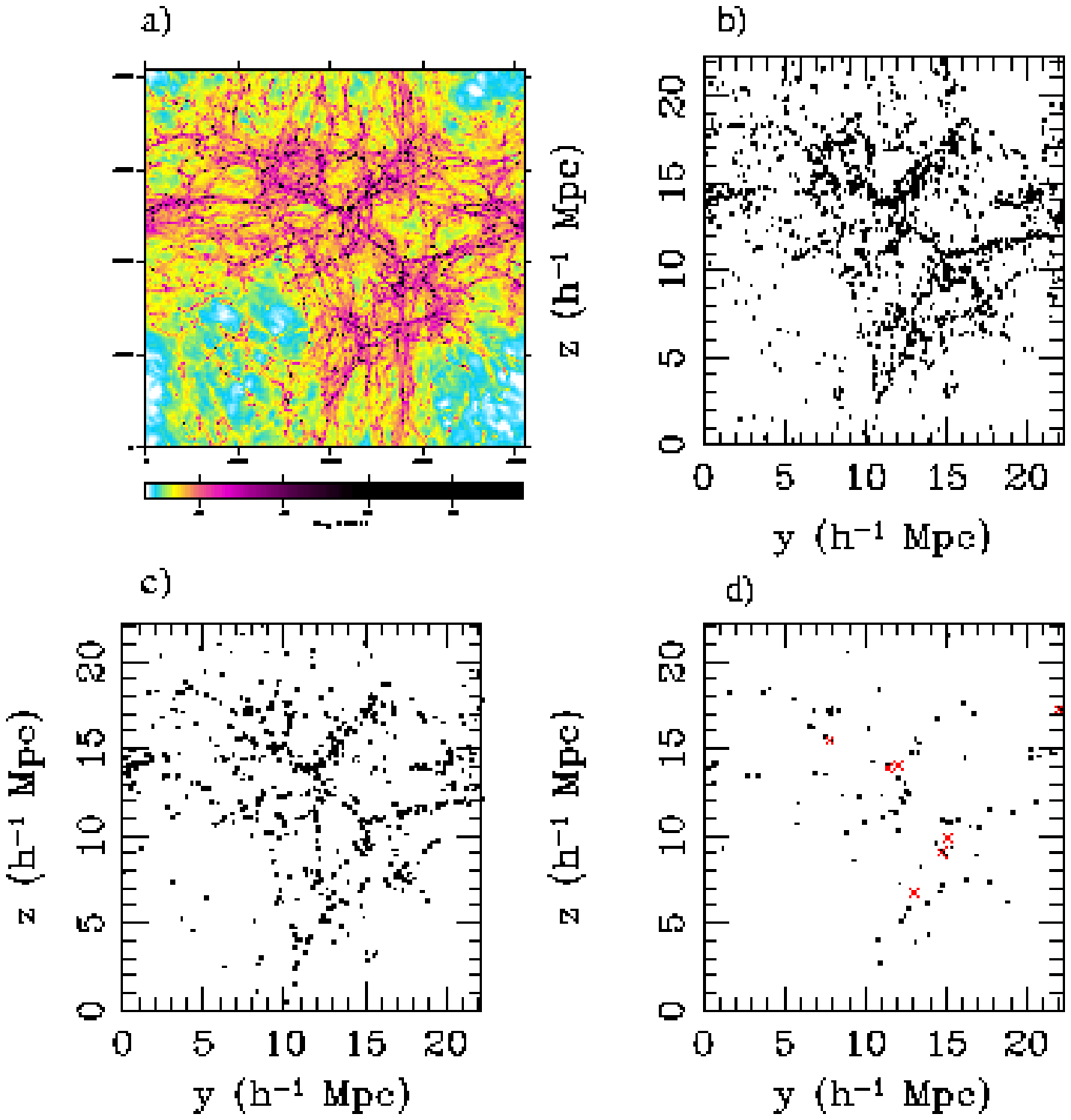}
\caption{({\it a}) Column density map of \HI\ in the $22.222~h^{-3}$~Mpc$^3$ volume projected along the $x$-axis
 on a  4096$^2$ pixel grid. Potential DLAs with  $N_{\HI}\;>\;10^{20.3}$~\cmsq\ appear black. 
({\it b})  Position of potential DLAs projected along the $x$-axis.  
({\it c})  Position of   the 651 galaxies  that have a baryonic mass $M_b$  larger than 
the resolution  $6.8\times 10^9$~\msun.
({\it d})  Position of the 100 most star-forming galaxies. The red crosses show the positions of the seven
most star-forming galaxies.
\label{cluster:fig:sim:DLAs}
}
\end{figure*}

We select DLAs from the simulations as follows.
We compute the \HI\ column density $(N_{\HI})$ from the gas density projected 
 onto a uniform grid with 4096$^2$ pixels, each 5.43~kpc comoving (or 2~kpc physical) in size,
 corresponding to the smoothing length. 
 Each gas particle is projected onto the grid  in correct proportions to the pixel(s) it subtends
given its smoothing length.  Since 
DLAs occur in dense regions, however, the smoothing lengths are typically 
equal to or smaller than the pixel size.
We first assume that the gas is optically thin, 
and then correct the column densities for the ionization background
using a self-shielding correction, as in \citet{KatzN_96a}.
The \HI\ column density projected along the $x$-axis is shown in  Figure~\ref{cluster:fig:sim:DLAs}({\it a}). 
A pixel is selected as a DLA from the column density map if   $N_{\HI}$ is greater than $10^{20.3}$\cmsq.
There are approximately 115,000 pixels that  meet this criterion, shown in Figure~\ref{cluster:fig:sim:DLAs}({\it b}).
We assume that each such pixel   is a potential DLA.
Figure~\ref{cluster:fig:sim:DLAs}({\it c})  shows  the  positions of the 651 galaxies that have a baryonic mass $M_b$  larger than 
the resolution  $6.8\times 10^9$ \msun.
Figure~\ref{cluster:fig:sim:DLAs}({\it d}) shows the positions of the 100 galaxies with the highest star formation rate,
and the positions of the simulated LBGs as red crosses. 
From Figure~\ref{cluster:fig:sim:DLAs}, one can already see that the galaxies and the DLAs are correlated.


\begin{figure*}
\plotone{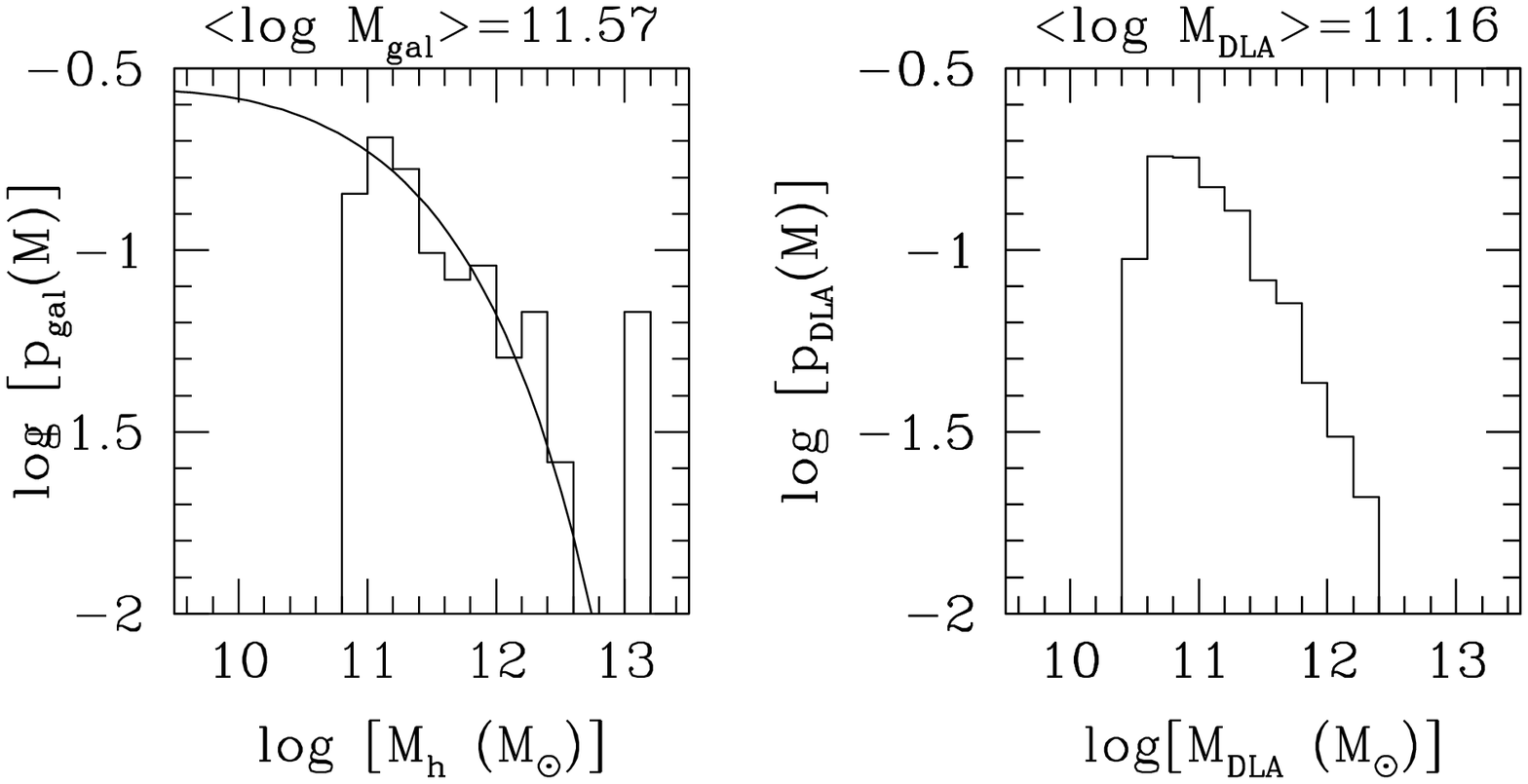}
\caption{{\it Left:} Halo mass (DM~$+$~baryons) probability function $p(M)$ 
 of all the SKID-identified galaxies with baryonic masses   larger than 
the resolution  $6.8\times 10^9$~\msun, corresponding to 64 SPH particles.
For comparison, the curve is the dark matter mass function from the extended Press-Schechter formalism of \citet{MoH_02a}
   in this cosmology, scaled arbitrarily (i.e.,  not a fit).
{\it Right:} DLA dark matter mass distribution, $p_{\rm DLA}(M)$ ($\propto \dNdzDLA$).
 This was found by matching the 2-D
DLA positions  with the nearest resolved halo. The shape of the distribution will not be constrained
by the DLA-galaxy cross-correlation, but its first moment ($\langle\log M\rangle$) will be.
\label{fig:distribution}}
\end{figure*}

The left panel of Figure~\ref{fig:distribution} shows the mass probability distribution of all the resolved galaxies.
The line shows the halo mass distribution obtained from the Press-Schechter (PS) formalism \citep{MoH_02a}.
The mean mass (logarithmic average) for all the 651 galaxies is shown ($\langle\log M_h (\msun)\rangle\;=\massLBGlog$).
The right panel of Figure~\ref{fig:distribution} shows  the DLA halo mass distribution. 
The halo mass of a given DLA was obtained 
by matching the projected DLA positions (2-D) with those of the resolved halos.
 The projected distance distribution (between halos and DLAs)
  peaks at 8~kpc, with a tail  to $\sim$20~kpc (physical units; see also Gardner et al. 2001),
and there is very little ambiguity in identifying the halo of a DLA.
 Practically all the DLAs reside in  halos with more than 64 dark matter particles.
Note that, at $z=0$, the DLA  distribution 
appears to be broadly peaked at around  $V_{\rm rot}=110$~\kms\ 
 \citep{ZwaanM_05a} and is even broader
with respect to luminosity \citep{RosenbergJ_03a}.


As mentioned, the purpose of this paper is to show 
that  cross-correlation techniques will  uniquely constrain the mean of this distribution, 
but   will not constrain  its shape. We refer the reader
to \citet{GardnerJ_01a} and \citet{NagamineK_04a} for a detailed discussion of the DLA
halo mass distribution in numerical simulations. Typically, in order to match the observed DLA statistics,
 they  require  an extrapolation of the DLA mass function below the mass
resolution.
Here we make no attempt to include halos smaller than our resolution since it would require
putting in the appropriate cross-correlation signal by hand for halos smaller than our resolution.

\section{Correlation functions in hierarchical models}
\label{section:clustering}

In this section, we describe the fundamental clustering relations necessary to understand
how one can determine the halo mass of DLAs.

A widely used statistic to measure the clustering of galaxies is the galaxy autocorrelation function $\xi_{\rm gg}(r)$.
Similarly, one can define the cross-correlation $\xi_{\rm dg}$
between DLAs and galaxies from the conditional probability of finding a galaxy in a volume d$V$ at a
distance $ r=|\mathbf r_1-\mathbf r_o|$, given that there is a DLA at $\mathbf r_o$,
\begin{equation}
 P(LBG|DLA)	=\overline n_u (1+\xi_{\rm dg}(r)) \mathrm d V, \label{eq:cross}
\end{equation}
where $n_u$ is the unconditional background galaxy density, i.e., the density when $\xi=0$.

At a given redshift, the autocorrelation  and cross-correlation  functions are related to the
dark matter correlation function $ \xi_{\rm DM}$ through the mean bias $\overline b(M)$ by 
\begin{eqnarray}
\xi_{\rm gg}(r)&=&\overline b^2(M_{\rm gal})\,      		\xi_{\rm DM}(r), \label{eq:biasCDM} \\
\xi_{\rm dg}(r)&=&\overline b(M_{\rm DLA})\, \overline b(M_{\rm gal})\, \xi_{\rm DM}(r), \label{eq:biascross}
\end{eqnarray}
where $M_{\rm gal}$ is the mean galaxy halo mass,   $M_{\rm DLA}$ is the mean DLA halo mass,
and $\overline b(M)$ is given by 
\begin{eqnarray}
\int_M^\infty p(M')\;b(M')\;\mathrm d M'\,, \label{eq:bias:integral}
\end{eqnarray}
where $p(M)$ is the halo mass probability distribution and $b(M)$ is the bias function, which can be 
computed using the extended  PS formalism \citep[e.g.,][]{MoH_93a,MoH_02a}.
Thus, from equations~\ref{eq:biasCDM}--\ref{eq:biascross},
  if both  $\xi_{\rm dg}$ and $\xi_{\rm gg}$ are power laws $(\xi\propto r^\gamma)$ with the same slope $\gamma$,
  the  amplitude ratio of the cross- to autocorrelation
   is a measurement of the bias ratio   $\overline b(M_{\rm DLA})/\overline b(M_{\rm gal})$,
from which one can infer  the halo masses $M_{\rm DLA}/M_{\rm gal}$. The details are presented in
 section~\ref{section:converttomass}.
Briefly, given that $b(M)$ is a monotonic increasing function of $M$,
if   $\xi_{\rm dg}$ is greater (smaller) than $\xi_{\rm gg}$, then 
 the halos   of DLAs are more (less) massive than those of the galaxies.

 In the remainder of this work, we   use only 
  projected correlation functions, $w(r_\theta)$,
where    $r_{\theta}=D_A(1+z)\theta$ in comoving Mpc, where
$D_A$ is the angular diameter distance.  This is  necessary since  (1) 
 the gas column density distribution is a 2-D quantity,
  and (2) this corresponds to the situation  when one relies on
   photometric redshifts \citep[e.g.,][BL04]{BoucheN_03a}.
Projected correlation functions   $w(r_\theta)$ is directly related to spatial correlation functions $\xi(r)$ if the 
selection function   is known.
Following \citet{PhillippsS_78a} and \citet{BudavariT_03a},
 the projected autocorrelation function, $w_{\rm gg}$,
  of galaxies with a redshift distribution ${\mathrm d N}/{\mathrm d z}$ is
\begin{eqnarray}
w_{\rm gg}(r_\theta)&=&\int_0^\infty d z \left(\frac{\mathrm d N}{\mathrm d z} \right)^2 g(z)^{-1} \times (f(z)
\theta)^{1-\gamma}\; r_{0}^\gamma \; H_\gamma \;,\nn\\  
&=& (r_\theta)^{1-\gamma}\; r_{\rm 0,gg}^\gamma \; H_\gamma 
\int_0^\infty d l \left(\frac{\mathrm d N}{\mathrm d l} \right)^2 \,, \label{eq:auto:result}  
\end{eqnarray}
where ${\mathrm d N}/{\mathrm d l}$ is the galaxy redshift distribution in physical units, 
$f(z)=D_A(1+z)$ is the comoving line-of-sight distance,
$g(z)=\mathrm d r/\mathrm d z=c/H(z)$, 
and  $H_\gamma=\Hgamma$ (see Appendix~\ref{appendix:cross}).
The projected cross-correlation $w_{\rm dg}$ 
between a given absorber at a given redshift
 and the galaxies (with a distribution ${\mathrm d N}/{\mathrm d z}$) is
\begin{eqnarray}
w_{\rm dg}(r_\theta)&=&\int \dNdl\ \xi(\sqrt{r_\theta^2+l^2}) \,\mathrm d l \label{eq:cross:result} \,.
\end{eqnarray}

For galaxies  distributed in 
a top-hat redshift distribution ${\mathrm d N}{\mathrm d l}$ of width $W_z$  [normalized such that $\int ({\mathrm d N}{\mathrm d l}) \mathrm d l=1$],
as  in the case here,
Equations~\ref{eq:auto:result} and \ref{eq:cross:result}
imply   that the amplitudes of both $w_{\rm dg}(r_\theta)$ and $w_{\rm dg}(r_\theta)$
are inversely proportional to  $W_z$ (see Appendix~\ref{appendix:cross} for the derivations):
\begin{eqnarray}
w_{\rm gg}(r_\theta)
&\simeq&(r_\theta)^{1-\gamma}\; r_{\rm 0,gg}^\gamma \; H_\gamma \times \left( \frac{1}{W_z}\right )^2 \; W_z \,, \\  
w_{\rm dg}(r_\theta)&\simeq&(r_\theta)^{1-\gamma}\; r_{\rm 0, dg}^\gamma \; H_\gamma \times \frac{1}{W_z}\,.
\end{eqnarray}
Therefore,  the ratio of the amplitudes of the two projected correlation functions
 $w_{\rm dg}$ to $w_{\rm gg}$  is simply  $(r_{\rm 0,dg}/r_{\rm 0,gg})^{\gamma}$, or
 the bias ratio  $\overline b(M_{\rm DLA})/\overline b(M_{\rm gal})$, from which we infer the mean DLA halo mass,
 regardless of the redshift distribution.
This is an important result for surveys that rely on photometric redshifts:
 the ratio of the {\it projected} correlations is a true measure of the bias ratio, 
 regardless of contamination or
uncertainty in the actual redshift distribution,
provided that the same galaxies are used for $w_{\rm dg}$ and $w_{\rm gg}$.

\section{Results}
\label{section:results}

In   section~\ref{section:results:cross}, we  quantify the amplitude of the DLA-galaxy  
cross-correlation relative to the galaxy-galaxy  autocorrelation in the SPH simulations.
 We show how to invert the cross-correlation results into a mass constraint in
 section~\ref{section:converttomass}.  We show that this method is independent of
 the galaxy sample that one uses (\S~\ref{section:samplevariation}).
  Finally, we compare these results
 to   observational results   in section~\ref{section:compareobservations}.

\subsection{DLA-Galaxy Cross-Correlation}
\label{section:results:cross}

\begin{figure}
\plotone{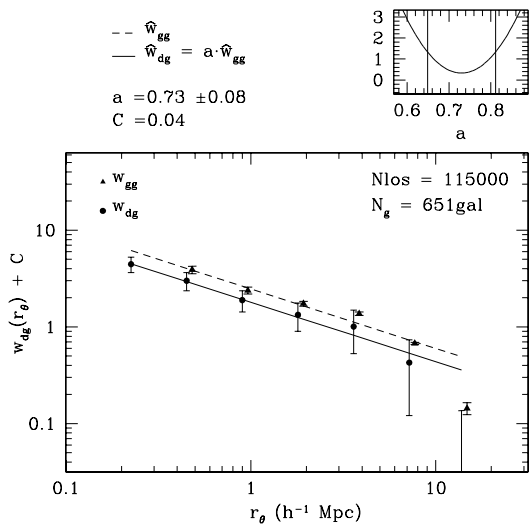}
 \caption{Filled circles show the projected DLA-galaxy  cross-correlation 
 $w_{\rm dg}(r_\theta)$ at $z=3$ in this $22.222~h^{-3}$~Mpc$^3$ simulation.
 The solid triangles show the projected autocorrelation $w_{\rm gg}(r_\theta)$
 (offset by 0.03dex in the $x$--axis for clarity).
 The full sample of 115,000 DLAs and 651 resolved galaxies was used.
 The amplitude ratio $a=\besta$, which is $w_{\rm dg}/w_{\rm gg}$, is found
 by fitting $w_{\rm dg}$ to the model $\hat{w}_{\rm dg}=a\; \hat{w}_{\rm gg}$,
  where $\hat{w}_{\rm gg}$ is the fit to the galaxy-galaxy autocorrelation  ({\it dashed line}).
 The small panel shows the $1\sigma$ range ({\it vertical lines}) of the $\chi^2$ distribution ({\it solid line}).
 An integral constraint of $C=0.04$ was used.}
 \label{fig:all:all}
 \end{figure}

The filled circles in Figure~\ref{fig:all:all} show the  DLA-galaxy cross-correlation $w_{\rm dg}$ 
using the entire sample of 115,000 DLAs and the 651 resolved galaxies.
We computed $w_{\rm dg}(r_\theta)$  with the  estimator 
\begin{eqnarray}
1+  w_{\rm dg}(r_\theta)&=& \left \langle  \frac{N_{\rm obs}(r_\theta)}{N_{\rm exp}(r_\theta)} \right \rangle, \label{eq:estimator}
\end{eqnarray}
where  $N_{\rm obs}(r_\theta)$ is  the observed number of galaxies   between $r_\theta-dr/2$ and $r_\theta+\mathrm d r/2$
from a DLA and  $N_{\rm exp}(r_\theta)$ is the expected number of  galaxies if they were uniformly distributed,
 i.e., $N_{\rm exp}(r_\theta)=2\pi r_\theta \Sigma_g \mathrm d r$ where 
 $\Sigma_g$  is the   galaxy surface density.
 $\langle\rangle$ denotes the   average   over the number of selected DLAs ($N_{\rm DLA}$).
In counting the pairs, we took into account the periodic boundary conditions of the simulations.


There are several reasons not to use other estimators such as the \citet[][LS]{LandyS_93a}  estimator.
First, we want to duplicate as closely as possible  the method (and estimator)  used in the 
observations of \citet{BoucheN_03a} and BL04. 
But, more importantly, the LS estimator
is symmetric under the exchange of the galaxies with the absorbers, whereas here and for the observations of BL04, the
symmetry is broken. This is due to the absorber redshift being well known, while galaxies have photometric
redshifts with larger uncertainties and, therefore, are distributed along the line of sight. This broken symmetry is also fundamental in
the derivation of equations~\ref{eq:auto:result}--\ref{eq:cross:result}. Had we used spectroscopic redshifts
and $\xi(r)$ instead of $w(r_\theta)$, the LS estimator would be superior.

Given that we use the galaxy surface density  $\Sigma_g$ to estimate the unconditional galaxy density 
(see eq.~\ref{eq:cross}), the integral of $\Sigma_g(1+w_{\rm dg})$ over the survey area $A$
will be equal to the total number of  galaxies, i.e., $\int_A  \Sigma_g (1+w_{\rm dg})\mathrm d A=N_g$. 
As a consequence, $\int  \Sigma_g\; w_{\rm dg}\;\mathrm d A=0$, and the correlation will be negative on the largest  scales,
 i.e., biased low.  This is the known  ``integral constraint''. In the case of our $22.222~h^{-2}$~Mpc$^2$ survey geometry,
  we estimated the integral constraint to be $C=0.04$, or 2\%\ of the cross-correlation
 strength at 1~\hMpc. We added $C$ to $w_{\rm dg}$ estimated from equation~\ref{eq:estimator}.
  
 The uncertainty  to $w_{\rm dg}$,   $\sigma_w$,  has two terms, the Poisson noise and the  clustering variance  
      \citep[see][and references therein; Appendix~\ref{appendix:errors}]{EisensteinD_03a}.
   In Appendix~\ref{appendix:errors}, we show that   $\sigma_w$
     is proportional to $1/\sqrt{N_{\rm DLA}}$ (eq.~\ref{cluster:eq:crossxi:variance}).

    There are several ways to compute $\sigma_w$ in practice. 
The proper way to compute $\sigma_w$
 would be to resample the  DLAs, since this would include 
   the  uncertainty due to the finite   the number of lines of sight.
However, this is valid for independent lines of sight, as in the case of an observational sample
(provided that $N_{\rm DLA}$ is large, say greater than $10$),
and will not be correct here given that   we have only one simulation
and   that we have to use the same galaxies for each simulated line of sight.
The uncertainty $\sigma_w$ must then  reflect   that we used only one realization of the large-scale structure.
For this reason, we elected to use the jackknife estimator
\citep{EfronB_82a}, i.e., by dividing the  $22.222h^{-2}$~Mpc$^2$ area into nine equal parts and  each time leaving one part
out. This will accurately reflect the uncertainty in $w_{dg}$ due to the one large-scale structure used,
but the signal-to-nois ratio (S/N$\equiv w_{\rm dg}/\sigma_w$) will not increase with $\sqrt{N_{\rm DLA}}$ as expected (Appendix B):
it will saturate   after a certain value of $N_{\rm DLA}$.
 We find that indeed the SNR saturates at $N_{\rm DLA}\simeq 40$
(not shown). This is a major difference
from observational samples, where   each field is independent. In that case,
  equation~\ref{cluster:eq:crossxi:variance} applies and  the S/N  is  proportional to $\sqrt{N_{\rm DLA}}$.

We computed  the full covariance matrix from
the $N_{\rm jack}=9$  realizations  as 
\begin{equation}
{\rm COV}_{ij}=\frac{N_{\rm jack}-1}{N_{\rm jack}}\sum_{k=1}^{N_{\rm jack}}  [w_k(r_{\theta_i})-\overline w(r_{\theta_i})]\cdot [w_k(r_{\theta_j})-\overline
w(r_{\theta_j})]\;, 
\label{eq:covariance}
\end{equation}
where $w_k$ is the $k$th measurement of the  cross-correlation and $\overline w$
is the average of the $N_{\rm jack}$ measurements of the cross-correlation.
The error bars in Figure~\ref{fig:all:all} show the diagonal elements of the covariance matrix,
i.e., $\sigma_w\equiv \sqrt{{\rm COV}_{ii}}$.

We computed the projected autocorrelation $w_{\rm gg}(r_\theta)$ of the same simulated galaxies used
for $w_{\rm dg}(r_\theta)$ in a similar manner.
 We used the estimator shown in equation~\ref{eq:estimator} to compute $w_{\rm gg}(r_\theta)$,
where $N_{\rm obs}(r)$ is now the number of galaxies between $r-\mathrm d r /2$ and $r+\mathrm d r /2$ from another galaxy.
The open triangles in Figure~\ref{fig:all:all} show the projected  autocorrelation $w_{\rm gg}(r_\theta)$ of the 651 galaxies.

We fitted  the galaxy autocorrelation with a power law model (${\hat w}_{\rm gg}=A_{\rm gg}r_\theta^\beta$)
 by minimizing $\chi^2\propto [\mathbf{w}-\mathbf{\hat w}]^T \mathrm{COV}^{-1}[\mathbf{w}-\mathbf{\hat w}]$,
where $\mathbf{w}$   and $\mathbf{\hat w}$ are the vector data and model, respectively,
and ${\rm COV}^{-1}$ is the inverse of the covariance matrix. 
We used single value decomposition (SVD) techniques to invert the covariance matrix,  ${\rm COV}$,
since it is singular and  the inversion is unstable  \citep[see discussion in][]{BernsteinG_94a}.

We then use that fit  as a template to constrain the amplitude of $w_{\rm dg}$, i.e.,
 \begin{equation}
 \hat w_{\rm dg}= a\times \hat w_{\rm gg},\label{eq:compareamplitude}
 \end{equation}
 where $a$ is  the amplitude ratio $A_{\rm dg}/A_{\rm gg}$ of the correlation functions.
This assumes that  the two correlation functions 
 have the same slope (see \S~\ref{section:clustering}). 
 This method also closely matches the method used by 
   BL04 (see \S~\ref{section:compareobservations} below)
 and   makes comparison to those observations straightforward.

   The solid line in Figure~\ref{fig:all:all} shows the fit to $w_{\rm dg}$ using  
    equation~\ref{eq:compareamplitude},    where the best amplitude $a$ is
\begin{equation}
a=\besta\label{eq:result:besta} \,.
\end{equation} 
The top panel shows the $\chi^2(a)$ distribution with the $1\sigma$ range. 
 In other words, 
the bias ratio $\overline b(M_{\rm DLA})/\overline b(M_{\rm gal})$ is $\besta$.
This can be converted into a correlation length   for $w_{\rm dg}$
of $a^{1/1.8}\simeq 0.84$ times that of the galaxy autocorrelation, i.e.,
$r_{\rm 0,dg}\simeq 0.84\; r_{\rm 0,gg}$.
 
Several authors \citep[e.g.,][and references therein]{BerlindA_02a,BerlindA_03a} have shown that
 the small scales ($r\langle1$~Mpc) of the correlation function
are  the scales sensitive  to variations in the halo occupation number.
At those scales, $\xi(r)$ is very susceptible to  galaxy pairs that are in the same halo.
Therefore, when we repeated our analysis with the six subsamples, we 
restricted ourselves to $r_\theta\,>\,1$~\hMpc.
In this case, for the full sample, we find the amplitude ratio to be $a=0.70\pm0.18$, 
in good agreement with equation~\ref{eq:result:besta}.

The reader should not use  these results (e.g., eq.~\ref{eq:result:besta}), obtained with 651 galaxies and 115,000  DLAs,
to scale the errors to smaller samples, because we use the same large-scale structure for all the 115,000
simulated DLAs.   As mentioned earlier, the large-scale structure  dominates the uncertainty at large $N_{\rm DLA}$,
and this is seen in the fact that the S/N saturates after $N_{\rm DLA}\simeq 40$.
We come back to this point at the end of \S~\ref{section:compareobservations}.

 \subsection{The Mass of DLA Halos from the Amplitude of $w_{\rm dg}$ }
\label{section:converttomass}
 
Equation~\ref{eq:result:besta}, i.e., the bias ratio
 $\overline b(M_{\rm DLA})/\overline b(M_{\rm gal})$,  can   be converted 
into a mean halo mass for DLAs if one knows the functional form of $b(M)$ and $M_{\rm gal}$.
One can use the PS  formalism \citep[e.g.,][]{MoH_02a} or the autocorrelation
of several galaxy subsamples to constrain the shape of $b(M)$.
 We   refer to these as the ``theoretical method'' and  as the ``empirical method,'' respectively.
 
\subsubsection{Theoretical Biases $b(M)$}

One can compute the theoretical biases   for any population (eq.~\ref{eq:bias:integral}) and predict
the bias ratio a priori if the mass probability distribution $p(M)$ is known.
Naturally, $p(M)$ is known   in our simulation (Fig.~\ref{fig:distribution}).
 We   show that the predicted bias ratio is well within the $1\sigma$ range
of our results (eq.~\ref{eq:result:besta}), demonstrating the reliability of the method.

\begin{figure*}
\plotone{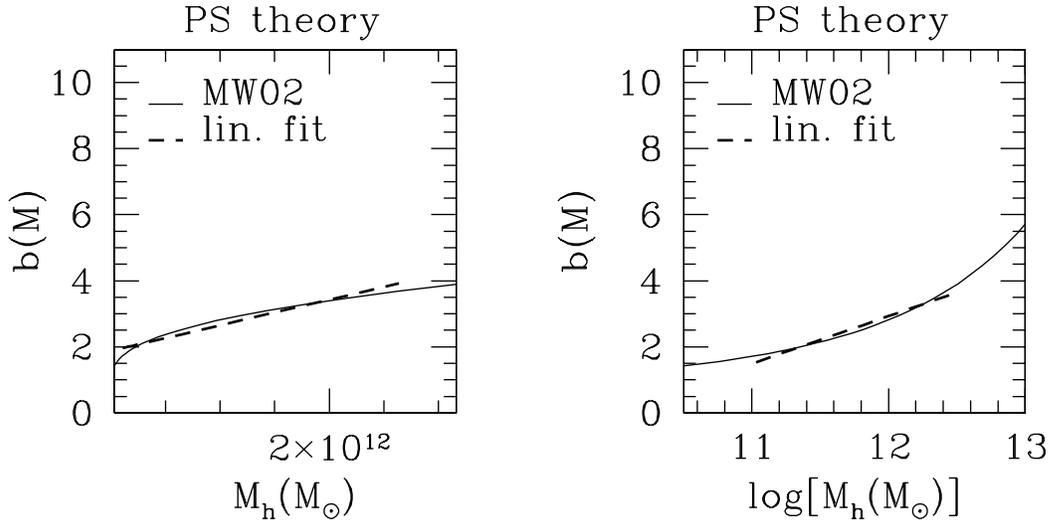}
\caption{{\it Left:} The $z=3$ bias $b(M)$ as a function of halos of mass $M$ from the extended Press-Schechter theory,
as in \citet{MoH_02a}.
{\it Right:} Same as left panel in $\log M$ space. 
In both panels, the dashed line is a linear fit to the curve over the mass range  $\log M\sim 11.5$---$12.5$.
\label{fig:bias:theory}}
\end{figure*}

 Given that galaxies and the DLAs
 actually lie in  halos of different masses, the theoretical biases are   found from equation~\ref{eq:bias:integral}, i.e.,
 \begin{eqnarray}
 \overline b_{\rm DLA}(> M)&=&\int_M^\infty p_{\rm DLA}(M')\;b(M')\;\mathrm d\log M' \label{eq:bias:DLA}\;, \\
 \overline b_{\rm gal}(> M)&=&\int_M^\infty p_{\rm gal}(M')\;b(M')\;\mathrm d\log M' \label{eq:bias:gals}\;, 
 \end{eqnarray}
 where $p(M)$ is the appropriate mass distribution [$p(M)\equiv \frac{\mathrm d n}{\mathrm d\log M}$
 normalized such that $\int p(M)\mathrm d\log M=1$]
  and $b(M)$ is the bias of halos of a given mass $M$.
The bias function $b(M)$ is also a function of redshift $z$, i.e., $b(M,z)$, and
  can be computed at a given $z$ from the extended PS  formalism  \citep[e.g.,][]{MoH_02a}.
 It  is shown in  Figgure~\ref{fig:bias:theory} for $z=3$
 on a linear-linear ({\it left}) and log-linear ({\it right}) plot.

The  mass distributions $p_{\rm DLA}$ and $p_{\rm gal}$  are shown in  
Figgure~\ref{fig:distribution}. 
Because $p(M)$ is bounded at some low-mass limit $M_{\rm min}$
(due to   limited resolution or to observational selection), the mean bias $\overline b$
of a given galaxy sample is defined by $\overline b_{gal}=\overline b(> M_{\rm min})$.

\begin{figure*}
\plotone{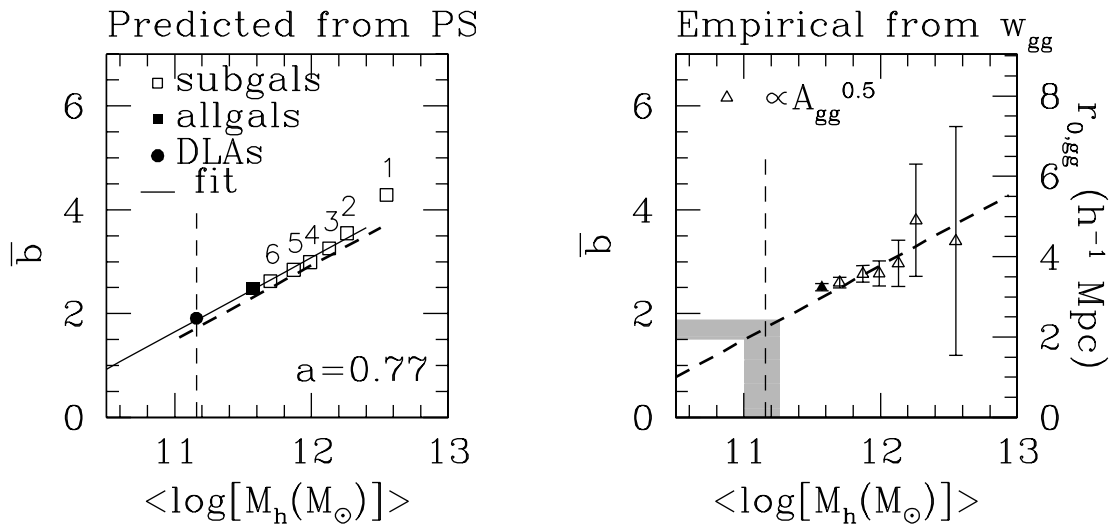}
\caption{{\it Left:} The symbols show the mean bias $\overline b$  as a function of the mean halo mass $\langle\log M_h\rangle$
for the DLAs ({\it circle}), full galaxy sample ({\it filled square}) and the subsamples ({\it open squares}), labeled 1--6.
The  bias is calculated using equations~\ref{eq:bias:DLA}--\ref{eq:bias:gals} and 
the distributions $p_{\rm DLA}$ and $p_{\rm gal}$ (shown in Fig.~\ref{cluster:fig:sim:mass}).
From the filled symbols, the predicted bias ratio is found to be \biasratiotrue.
  The solid line is a linear fit  to the points and is 
  5\%\ away from  to  the linear approximation (Fig.~\ref{fig:bias:theory}) shown by the thick dashed line.
{\it Right:} Same as left panel, but using the
empirical method (see text) instead of the mass distributions. The triangles show the mean bias
for the different galaxy samples using $\overline b \propto \sqrt{A_{gg}}$ (eq.~\ref{eq:biasCDM}) where
the normalization    was adjusted to match the amplitude of the dashed line.
The bias for the 651 galaxies is represented by the filled triangle.
The open triangles show $\overline b$ for the  subsamples. The right axis scale shows the
corresponding correlation length $r_0$.
For the DLAs, the shaded areas show the   measured
bias ratio $a=\besta$  and  
the corresponding halo mass range $\langle\log M_{\rm DLA} (\msun)\rangle\;=\massDLAlog$ 
using the mean mass of the 651 resolved galaxies: $\langle\log M_{\rm gal}(\msun)\rangle\;=\massLBGlog$.
In both panels, the vertical dashed line shows the ``true'' 
mean $\langle\log M_{\rm DLA}(\msun)\rangle=\massDLAlogtrue$ obtained from 
the DLA mass distribution $p_{\rm DLA}(M)$ shown in Figure~\ref{fig:distribution}. Both panels are for $z=3$.
\label{fig:bias:compare}}
\end{figure*}

The predicted biases $\overline b$  are   shown in the left panel of Figure~\ref{fig:bias:compare}.
The predicted biases $\overline b$ for the subsamples, the  651 galaxies, and the DLAs 
are represented by open squares, the filled square and the filled circle, respectively.
From  $\overline b$  for the 651 galaxies ({\it filled square}) and for  DLAs ({\it filled circle}),
the theoretical bias ratio $\overline b_{\rm DLA}/\overline b_{\rm gal}$ is
found to be \biasratiotrue,
very close to the bias ratio measured from the clustering of galaxies around the DLAs (eq.~\ref{eq:result:besta}).

 When the distributions $p(M)$ are not known, we can infer a mass ratio from
the bias ratio using the approximation   $b(M)=b_0+b_1\;\log M$,\footnote{One can use
 $b(M)=b_0'+b_1'\; M$ instead, and replace $\langle\log M\rangle$ by $\langle M\rangle$ in
the remaining of the discussion.} over a restricted mass range.
In each panel in  Figure~\ref{fig:bias:theory}, the dashed line shows
such a linear fit over the mass range $\log M\sim 11$---$12.5$.
Using this approximation, the mean bias $\overline b$ is given by
\begin{eqnarray}
\overline b(\> M_{\rm min})&=&\int_{M_{\rm min}}^\infty p(M')\;b(M')\;\mathrm d\log M'  \nn \\
&=&b_0+b_1\;\int_{M_{\rm min}}^\infty p(M')\; \log M'\;\mathrm d\log M' \nn \\
&=&b_0+b_1\;\langle\log M\rangle  \label{eq:bias:approx}\;,
\end{eqnarray} 
where $\langle\log M\rangle$  is the first  moment  of the distribution $p(M)$.
Thus, the mean bias for the galaxies and the DLAs
are  $\overline b_{\rm gal}=  b(\langle\log M_{\rm gal}\rangle)$, 
and  $\overline b_{\rm DLA}=  b(\langle\log M_{\rm DLA}\rangle)$, respectively,
 where $\langle\rangle$ denote the first moment of the appropriate mass distribution.

In Figure~\ref{fig:bias:compare} ({\it left})
the solid line is a linear fit
 to the theoretical biases $\overline b(\langle\log M\rangle)$, and  is 5\%\ away (in amplitude) from   
  the linear approximation (eq.~\ref{eq:bias:approx})
shown by the thick dashed line.
The vertical dashed line indicates the mean DLA halo mass $\langle\log M_{\rm DLA}\rangle$ that is found from
the first moment of the mass distribution in Figure~\ref{fig:distribution}.
This shows that using a linear approximation of $b(M)$ is  equivalent  to using the bias function $b(M)$ 
from \citet{MoH_02a}, provided that the DLA-galaxy mass ratio is not larger than a decade.
Indeed, the 5\%\ difference in amplitude
cancels out when taking the bias ratio.

\subsubsection{Empirical Method for $b(M)$}

To infer $\langle\log M_{\rm DLA}\rangle$ from  equation~\ref{eq:result:besta} or from  observations,
one needs to find the coefficients $b_0$  and $b_1$. To do so,
one can either use the PS formalism \citep{MoH_02a} 
or use the fact that $\overline b$ is proportional to  $\sqrt{A_{\rm gg}}$ (eq.~\ref{eq:biasCDM}), 
where  $A_{\rm gg}$ is measured for each of the galaxy subsamples covering the mass range $\log M\sim 11.5$---$12.5$.
Figure~\ref{fig:bias:compare} ({\it right}) illustrates this point. 
The thick dashed line is again
the linear approximation shown in Figure~\ref{fig:bias:theory}.
The  open (filled) triangles show the mean biases $\overline b$  of the subsamples (full sample) assuming
that  $\overline b\propto \sqrt{A_{\rm gg}}$ (eq.~\ref{eq:biasCDM}).
The normalization is set to match the dashed line, and
 is not relevant, since we measure a ratio of two biases.
  This shows that one can use either  the PS formalism \citep{MoH_02a} 
or use  $\sqrt{A_{\rm gg}}$ to find the coefficients $b_0$  and $b_1$.

 In the case where  the autocorrelation length $r_{\rm 0,gg}$
 has been determined, one can use 
 the right $y$--axis scale of Figure~\ref{fig:bias:compare} the infer the DLA halo mass
 from the measured bias ratio.

\subsubsection{The Mean DLA Halo Mass}

To   actually determine $\langle\log M_{\rm DLA}\rangle$ from our cross-correlation result (eq.~\ref{eq:result:besta}), we used
(1) the linear approximation to the PS bias (Fig.~\ref{fig:bias:compare}, {\it thick dashed line}), and
(2)  $\langle\log M_{\rm gal}\rangle=\massLBGlog$ for  the 651 galaxies.
We infer a mean DLA halo mass of $\langle\log M_{\rm DLA}\rangle=\massDLAlog$,
shown by the vertical shaded area on the right panel of Figure~\ref{fig:bias:compare}.
Our cross-correlation result (eq.~\ref{eq:result:besta}) is shown by the horizontal shaded area.
The ``true'' DLA mass derived from $p_{\rm DLA}$ (Fig.~\ref{fig:distribution})
and equation~\ref{eq:bias:DLA}
is shown by the vertical dashed line at $\log M_{\rm DLA}=\massDLAlogtrue$.
Similarly, using fits to $b(M)$ in linear space (Fig.~\ref{fig:bias:theory}, {\it left}),
 we infer $\langle M_{\rm DLA}\rangle=\massDLAall$~\msun, close to ``true'' mean  $1/N_{\rm DLA}\sum_i M_{\rm DLA,i}=\massDLAtrue$~\msun.

 In summary,  the amplitude of $w_{\rm dg}$ relative to $w_{\rm gg}$, $a=\besta$ (eq.~\ref{eq:result:besta}), 
 measured in this simulation implies that  DLAs 
 have halos of (logarithmically) averaged mass
\begin{equation}
\langle\log M_{\rm DLA} (\msun)\rangle\;=\massDLAlog\;,    \label{eq:result:massDLA}   
 \end{equation}
close to the true \massDLAlogtrue.
This shows that the cross-correlation technique uniquely constrain the mean of the halo mass distribution,
  despite the fact that DLAs occupy a range of halo masses and
some halos contain multiple galaxies and multiple DLA systems.
 In \S~\ref{section:samplevariation}, we   show that the technique is reliable in the 
sense that it will lead to the same answer regardless of the galaxy sample used.

From the right panel of Figure~\ref{fig:bias:compare}, we can now predict the cross-correlation strength for
 real $z=3$ LBGs, which have a correlation length of $r_{\rm 0,gg} \simeq4$~Mpc \citep[e.g., ASSP03;][]{AdelbergerK_04a},
corresponding to a halo  mass of $M_{\rm h}\simeq 10^{12}$~\msun. From the figure,  one  expects
that the correlation ratio or the bias ratio is $\sim 1.75/3=0.58$, and thus
the DLA-LBG cross-correlation would have a correlation length $r_{\rm 0,dg}= 4\times(0.58)^{1/1.6}\simeq \rnotLBG$~Mpc.

Potential systematic errors  include the few massive halos ($M_h > 10^{13}$~\msun) that are missed
 due to the limited volume ($22.222~h^{-3}$~Mpc$^3$) of our simulation.
However, since DLAs are cross section selected these few massive halos are too scarce to change
the mean $\langle\log M_{\rm DLA}\rangle$ of the DLA mass probability distribution (Fig.~\ref{fig:distribution}, {\it right}).
Naturally, if there were such massive halos in our simulations, the amplitude of the cross-correlation
would be different. We address this point in a general way in \S~\ref{section:samplevariation} and
show that the derived $\langle\log M_{\rm DLA}\rangle$ is independent of the galaxy sample used.
 
Our treatment of feedback    is limited to energy injection of supernovae, and thus
does not treat phenomena like winds.  \citet{NagamineK_04a} included winds in similar simulations and
showed that the DLA abundance  decreases with increasing wind strength, but
the mean DLA halo mass  will be shifted towards higher mass in the presence of winds.
 \citet{NagamineK_04a} also showed that the DLA abundance (extrapolated to $10^8$~\msun, i.e.,
 below the resolution limit, using the PS formalism)
 also decreases with increasing resolution, but again, 
the mean DLA halo mass will be shifted toward  higher mass in higher resolution runs.

Given that (1)   a   fraction of   DLAs are expected to arise in   halos   below our mass resolution of
  $M_h\ga 5.2\times 10^{10}$~\msun\ and  
(2) our total DLA abundance extrapolated to $10^8$~\msun, as in \citet{NagamineK_04a},  
over-predicts the observed DLA abundance, 
 equation~\ref{eq:result:massDLA}  is an upper limit. 
Furthermore, given  the results of \citet{NagamineK_04a} showed
that both winds and better resolution increase  the mean DLA halo mass,
 we conclude that a simulation with SNe winds and with
a better mass resolution  would  lower our mean DLA mass.
Reading from Figure~5 in \citet{NagamineK_04a}, we estimate that $\langle\log M_{\rm DLA}\rangle$ is $\sim10.6$
in their high-resolution run with strong winds, or a factor of $\sim5$ smaller than here.
Thus,  equation~\ref{eq:result:massDLA}  is an upper limit. 

\subsection{The Cross-Correlation Is Independent of the Galaxy Sample}
\label{section:samplevariation}
   
   From  equations~\ref{eq:biasCDM} and \ref{eq:biascross}, we expect 
the relative amplitude $a$ to vary as a function of the
halo mass of the galaxy  sample $M_{\rm h}$.
We therefore performed the same cross-correlation calculations   for each of the six subsamples
presented in \S~\ref{section:simulation} (see also Fig.~\ref{cluster:fig:sim:mass}),
and ask the question, is the inferred $\langle M_{\rm DLA}\rangle$ the same in each case?
We restricted ourselves to scales
$r_\theta \,>\,1$~\hMpc\ (from the discussion in \S~\ref{section:results:cross}).   

\begin{figure}
\epsscale{0.85}
\plotone{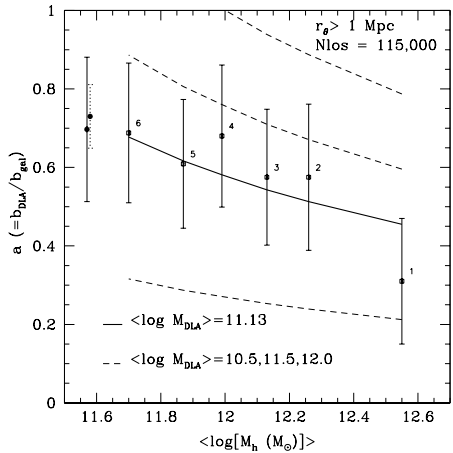}
\caption{DLA-LBG cross- to autocorrelation amplitude ratio,
 $a$ (defined in eq.~\ref{eq:compareamplitude}),
as a function of the mean galaxy halo mass $\langle\log M_h\rangle$ using the full DLA sample of $\sim115,000$ lines of sight.
Note that $a$ is also  $\overline b(M_{\rm DLA})/\overline b(M_{\rm gal})$.
Each of the subsamples (see Fig.~\ref{cluster:fig:sim:mass}) is labeled  1--6.
The filled circle with solid error bars shows  $a$ for the entire sample of 651 galaxies and 115,000 DLAs.
The filled circle with dotted error bars (offset along the $x$-axis) shows $a$ fitted over
all scales  for the entire sample of 651 galaxies and 115,000 DLAs 
from which we infer a mean DLA halo mass of $\langle\log M_{\rm DLA}\rangle=\massDLAlog$.
For this mass,  the expected amplitude ratio $a$  for the six subsamples 
 is shown by the solid  line. The expected $a$ follows closely the values found for the subsamples, showing
 that one will get the same DLA halo mass for any galaxy subsample, as long as the same galaxies are used
 for $w_{\rm gg}$ and $w_{\rm dg}$, i.e.,  that the method is reliable and self-consistent. For comparison,
the dashed lines show the expected $a$ if the mean DLA halo mass $\langle\log M_{\rm DLA}\rangle$
 were 10.5, 11.5, and 12 (from bottom to top) instead.
}
\label{fig:compareB}
\end{figure}

Figure~\ref{fig:compareB}  shows  the measured amplitude or bias ratio $a$   for each of the subsamples.
 The amplitude ratio $a$  for the subsamples (full sample) is 
  represented by the open squares (filled circle) with solid error bars. 
The filled circle with dotted error bars 
represents  the full sample shown in Figure~\ref{fig:all:all}   from which we inferred 
$a=\besta$ and $\langle \log M_{\rm DLA}\rangle=\massDLAlog$.
As expected,   $a$ increases with larger subsamples, or  with
 decreasing     galaxy halo mass $M_{\rm h}$.


For the method to be self-consistent,
the derived DLA halo mass $\langle\log M_{\rm DLA}\rangle$ should be the same for all the subsamples.
Given equation~\ref{eq:bias:approx},   $\langle\log M_{\rm DLA}\rangle=11.13$ determined in \S~\ref{section:converttomass}, and a
mean galaxy halo mass $\langle\log M_h\rangle$, we can predict the bias ratio $a$.
The solid line in Figure~\ref{fig:compareB} represents this prediction.
One sees that the   measured  bias ratio for the subsamples ({\it open squares}) follow the 
expected bias ratio ({\it solid line}).
 For comparison, the  dashed lines show the expected amplitude ratios $a$ if DLAs were in halos
of mean mass $\langle\log M_{\rm DLA}\rangle=10.5$, 11.5, and 12 (from bottom to top)
instead of the inferred $\langle\log M_{\rm DLA}\rangle=11.13$.

We conclude that the method is reliable and self-consistent, i.e., the mass $\langle\log M_{\rm DLA}\rangle$
 is independent of the galaxy sample used,
 that the clustering statistics of DLAs with galaxies can be used to infer
their mass, and that large observational samples will shed new light on their nature.
A direct observational measure of the relative amplitude $a$ (J. Cooke 2005, private communication), will show whether or not 
DLAs are massive  disks ($10^{12}~\msun$) as proposed by \citet{WolfeA_86a,WolfeA_95a} and \citet{ProchaskaJ_97b}.

\subsection{Comparison to  Observations}
 \label{section:compareobservations}

In this section, 
we first briefly review past and recent observations of clustering between
galaxies and DLAs (\S~\ref{section:currentobservations}).
We then (\S~\ref{section:smallsample}) focus on comparing the simulated DLA-LBG cross-correlation
to the observational results of  BL04, in a meaningful way, i.e., 
with a sample of similar size.

\subsubsection{Observations of the DLA-Galaxy Cross-Correlation at $z=3$}
 \label{section:currentobservations}
 
Early  attempts  to detect diffuse \Ly\ emission from DLAs at $z>2$ using
 deep	narrow band imaging \citep{LowenthalJ_95a} 
   did not reveal the absorber  but unveiled a few companion \Ly\ emitters
   \citep{LowenthalJ_91a}, hinting at the clustering of galaxies around DLAs. 
This prompted \citet{WolfeA_93a} to calculate the two-point correlation function
at $\langle z\rangle=2.6$ and to conclude that, indeed, \Ly\ emitters are clustered near DLAs
at the  $99$\%\ or greater confidence level.
Some recent \Ly\ searches have succeeded in unveiling the absorber \citep[e.g.,][]{FynboJ_99a}.

 \citet{FrancisP_93a} reported the discovery of super-clustering of sub-DLAs at $z\sim 2.4$ and $2.9$:
a total of four \HI\ clouds are seen in a QSO pair separated by 8\arcmin, each being at the same velocity.
Recent results from narrow-band imaging of the Francis \& Hewett field show that spectroscopically confirmed
\lya\ emitters are clustered at the redshift of the strongest \HI\ cloud at $z=2.9$ ($\log N_{\HI}=20.9$) towards Q2138-4427 
\citep{FynboJ_03a}. \citet{RocheN_00a} identified eight
	\Ly\ emitting galaxies near the DLA at $z=2.3$ towards PHL 
	957 in addition to the previously discovered Coup Fourr\'e galaxy 
	\citep{LowenthalJ_91a}, implying the presence of a group, 
	filament, or proto-cluster associated with the DLA.   
Other evidence of clustering includes the work of \citet{EllisonS_01a}, who found
 that the DLA at $z_{\rm abs}=3.37$ towards
 Q0201+1120 is part of a concentration
 of matter that includes at least four galaxies (including the DLA) over
 transverse scales greater than $5~h^{-1}$~Mpc, and of
 \citet{DodoricoV_02a} who showed  that out of 10 DLAs in QSO pairs, five are matching
 systems within 1000~\kms.
 They concluded that this result indicates a highly significant over-density of
 strong  absorption systems over separation lengths from $\sim 1$ to $8~h^{-1}$~Mpc.

\citet{GawiserE_01a} studied the cross-correlation of LBGs around one $z\sim 4$ DLA.
 Probably due to the high redshift of their DLA,  \citet{GawiserE_01a} found that $w_{\rm dg}(r_\theta)$
 is consistent with 0, i.e., they found that the distribution of the eight galaxies in that field (with spectroscopic redshifts)
 is indistinguishable from a random distribution.
 Their data did not allow them to   put limits on the amplitude of $w_{\rm dg}$.

Recently, ASSP03 found a lack of galaxies near  four DLAs and concluded
that the DLA-LBG cross-correlation is significantly weaker than the LBG-LBG autocorrelation
at the 90\%\ confidence level.
They found two LBGs within
$r_\theta=5.7$~\hMpc\ and within $W_z\langle0.0125$ ($\sim8$~\hMpc) whereas $\sim6$ were expected 
if the cross-correlation has the same amplitude as the galaxy autocorrelation. 
Because of the field of view available,  both \citet{GawiserE_01a} and ASSP03
 were not sensitive to scales larger than $r_\theta\sim5~h^{-1}$~Mpc,
which is important  since
the relevant scales to measure   the DLA-LBG cross-correlation extend
up to $r_\theta\sim10~h^{-1}$~Mpc.

However, the results of ASSP03 can be used to put an upper 
limit on $w_{\rm dg}/w_{\rm gg}$ through the following steps:
First, note that
the two galaxies   (in $N_{\rm DLAs}=4$ fields) observed by ASSP03 give 
\begin{eqnarray}
\langle N_{\rm obs}\rangle=2/4=0.5=\langle N_{\rm exp}\rangle(1+\overline \xi_{\rm dg})\,,\label{eq:adelberger}
\end{eqnarray}
 and the six galaxies expected if $\overline \xi_{\rm dg}=\overline \xi_{\rm gg}$ give
  $\langle N_{\rm exp}\rangle(1+\overline \xi_{\rm gg})=6/4$,
   where $\overline \xi$ is the volume average of the correlation function.
Second, for the LBG autocorrelation published in
ASSP03, we find $\overline \xi_{\rm gg}\simeq 1.1$, averaged over a sphere centered on the DLAs with an effective radius of $\sim6$~\hMpc, i.e., with
the same volume as the cylindrical cell used by ASSP03.
Thus, the   expected number of galaxies per DLA field  is $\langle N_{\rm exp}\rangle=0.68$  if $\overline \xi_{\rm dg}=0$,
and the total number of galaxies is $4\times \langle N_{\rm exp}\rangle=2.85$. 
Clearly their measurement of two galaxies is consistent with no cross-correlation.
From equation~\ref{eq:adelberger}, we can infer that $1+\overline \xi_{\rm dg}=0.7$  using $\langle N_{\rm exp}\rangle=0.68$.
Third, the uncertainty to $\overline \xi_{\rm dg}$, $\sigma_\xi$, can be estimated
using  the results shown in Appendix~\ref{appendix:errors}. 
The variance $V(\xi)=\sigma_\xi^2$ is made of two terms,
 the shot noise variance $V_{\rm sn}$ and the clustering
 variance $V_{\rm cl}$.
The shot noise variance to $\langle N_{\rm obs}\rangle$ is $V(\langle N_{\rm obs}\rangle)_{\rm sn}=\langle N_{\rm obs}\rangle=0.5$ 
(eq.~\ref{cluster:eq:crossvariance:sn}).
The two-point clustering variance (eq.~\ref{cluster:eq:crossvariance:2pt:peebles}) is simply
$\overline N^2(A\overline \xi_{\rm gg})=\overline N^2(2.50)$ where $A=J_2/K_1=2.28$.
The three-point clustering variance (eq.~\ref{cluster:eq:crossvariance:3pt:peebles}) is $\sim 0$ since $\xi_{\rm dg}\simeq 0$.
Finally, from equation~\ref{cluster:eq:crossxi:variance}, 
$\sigma_\xi=\frac{1}{\sqrt{4}}\frac{1}{\sqrt{2.85/4}}\sqrt{(1+\frac{2.5}{0.7})}\sqrt{0.7}\simeq 1.06$, 
and a 1-$\sigma$ (2-$\sigma$) upper limit to $\overline \xi_{\rm dg}$ is 
$\overline \xi_{\rm dg}+1(2)\sigma_\xi=-0.3+1.06(2.12)=0.76(1.82)$.
Since $\overline \xi_{\rm gg}=1.1$, the 1-$\sigma$ (2-$\sigma$) upper limit to the amplitude ratio is
$\overline \xi_{\rm dg}/\overline \xi_{\rm gg}\;\la\;0.70(1.65)$, respectively.
 This rough calculation is quite consistent with Adelberger's results where it was found
that   $\xi_{\rm dg}<\xi_{\rm gg}$  at the  90\%\  confidence level using Monte Carlo simulations.

Given that the relevant scales to measure   the DLA-LBG cross-correlation extend
up to $r_\theta\sim 10$~\hMpc, \citet{BoucheN_03a} 
were able to first detect and measure a DLA-LBG cross-correlation signal  (BL04)
using the wide-field  ($0.35$~deg$^2$ or $\sim 40^2$~Mpc$^2$ comoving at redshift $z=3$)
imager MOSAIC on the Kitt Peak 4m telescope.
\citet{BoucheN_03a} showed that there  was  an over-density of LBGs by a factor of $\sim3$ (with 95\%\ confidence)
 around the $z_{\rm abs}\simeq 3$ DLA towards the quasar APM 08279+5255 ($z_{\rm em}=3.91$) 
on scales $2.5<r_\theta<5$~\hMpc.
Extending the results of \citet{BoucheN_03a} to three  $z\sim 3$ DLA  fields,
 BL04   probed the DLA-LBG cross-correlation on scales $r_\theta\sim5$--$20$~\hMpc\
and found
 (1) a  DLA-LBG   cross-correlation with  
 a relative amplitude $w_{\rm dg}=(1.62\pm1.32)\, w_{\rm gg}$
   that is greater than zero at the $\sim 95$\% confidence level,
 and (2) that $w_{\rm dg}$ is most significant on scales 5--10~\hMpc.
 In other words, DLAs are clustered with LBGs, but unfortunately
 the sample size did not allow BL04 to test whether $a$ is greater or smaller than 1.
  Soon, the ongoing survey of 9 $z\simeq 3$ DLAs of
\citet{CookeJ_05a} will triple the sample of BL04.

In a slightly different context, \citet{BoucheN_04a} applied successfully the technique
presented here to  212 $z\simeq0.5$ \MgII\ systems (of which 50\%\ are expected to be DLAs)
 using luminous red galaxies (LRGs) in the Sloan Digital Sky Survey Data Release 1.
 They found that the \MgII--LRG
cross-correlation has an amplitude \Arelcorr\ times that of the LRG--LRG
autocorrelation, over comoving scales up to $r_\theta=13$~\hMpc. Since LRGs have halo-masses greater than
\LRGminmass\,\msun\ for $M_R\!\la\!-21$, this relative amplitude
implies that the \MgII\ host galaxies have halo masses greater than
\LRGmassrangemin.
These results show how powerful  the cross-correlation technique is.

To summarize the current observational situation on the $z=3$ DLA-LBG cross-correlation,
ASSP03  finds that    the amplitude ratio is
$\overline \xi_{\rm dg}/\overline \xi_{\rm gg}\;\la\;0.70$,  and 
BL04 finds that 
$\overline \xi_{\rm dg}/\overline \xi_{\rm gg}\;\ga\;0.30$,  both at the 1-$\sigma$ level.
Using Monte Carlo simulations, 
ASSP03  finds   $\overline \xi_{\rm dg}/\overline \xi_{\rm gg}\;<\;1.0$,  at the 90\%\ confidence level, and
BL04 finds  $\overline \xi_{\rm dg}/\overline \xi_{\rm gg}\;>\;0.0$,  at the 95\%\ confidence level.
The DLA halo mass range allowed by these observations is still large: it covers 
$\log M_{\rm DLA}\sim {10}$--${12}$~\msun.\footnote{After this paper's
submission, we learned that P. Monaco et al.(2005, private communication) constrained  the halo mass
of a few individual DLAs to be around $5\times 10^{11}$ \msun. Their mass estimates come
   from the emission-absorption redshift difference as a proxy for a rotation curve.}

\subsubsection{Simulation of Present Observations: $w_{\rm dg}$ with Small Samples}
\label{section:smallsample}

There are many significant differences between the observational sample of BL04
and the present simulated one.
First, the shape of the volume is very different:
the survey volume of BL04 is $40\times40\times100~h^{-3}$~Mpc$^3$ (comoving), while these simulations 
are $22.222~h^{-1}$~Mpc (comoving) on a side. Given that the survey of BL04 contains  about 80--120 LBGs per field,
their observed LBG number density corresponds to about  seven galaxies per  $22.222~h^{-3}$~Mpc$^3$. 
Naturally, seven galaxies are not a fair sample of the LBG luminosity function.
This is an inherent problem due to the size of the simulation, rendering the comparison between
the observed and the simulated cross-correlation difficult. 
 Second, as mentioned in \S~\ref{section:simulation},
 the simulated LBGs are selected according to their SFR, while the observed LBGs are color selected.
Third, the same galaxies are used for every simulated line of sight.
These differences limit our ability to perform a direct comparison to observations.

 With these caveats in mind, we  can repeat our analysis of section~\ref{section:converttomass} 
in the limit of small $N_{\rm DLA}$ and with similar galaxy number densities.
 Because, to first order, $\sigma_w\propto (\sqrt{N_{\rm DLA}N_{\rm gal}})^{-1}$ (eq.~\ref{cluster:eq:crossxi:variance}),
 a sample made of 10 DLAs and 25 galaxies per $22.222^2~h^{-2}$~Mpc$^{2}$ ``field'' 
 is expected to have similar errors to the sample of BL04 made
of three DLAs and 100 galaxies per $40^2~h^{-2}$~Mpc$^{2}$ field.
As for the full sample, we restricted ourselves to scales $r_\theta\;>\;1$~\hMpc,
which also corresponds   to the most relevant scales  5--10~$h^{-1}$~Mpc of the
 observations of BL04.
We find that  the relative amplitude of the cross-correlation with 10 lines of sight
  and 25 galaxies  is $a=0.77\pm0.53$,   whereas 
 BL04 found $a=1.62\pm1.32$, i.e.,  
both with the same S/N.

This confirms the results of BL04. More importantly, one can  now use the
result for this sample made of  10 DLAs and
25 galaxies (with a surface density $\Sigma_{\rm g}\sim 0.05$ Mpc$^{-2}$) as
a benchmark to predict the S/N for  the larger samples of future observations, given
that the S/N will be proportional to 
 $\sqrt{N_{\rm gal}N_{\rm DLA}}$ (eq.~\ref{cluster:eq:crossxi:variance}).

 \section{Conclusions}
 \label{section:conclusions}
 
 Motivated by the fact that (1) the amplitude of the cross-correlation is a measurement of the mean DLA halo
mass and (2) observational constraints 
(\citeauthor{GawiserE_01a}~\citeyear{GawiserE_01a}; ASSP03; BL04; J. Cooke et al. 2005, private communication)
are reaching a turning point and the DLA halo masses are starting to be constrained,
we tested the cross-correlation technique using
 TreeSPH cosmological simulations.
 The method uses the ratio of the cross-correlation
between DLAs and high-redshift galaxies to the autocorrelation
of the galaxies themselves, which is (in linear theory) the ratio of their bias factor,
 to infer the mean DLA halo mass.

In a  TreeSPH simulation \citep{KatzN_96b}  parallelized by \citet{DaveR_97b}
with $128^3$ particles  in a volume 22.222$^3$~$h^{-3}$~Mpc$^3$ (comoving), 
we find the following:
\begin{enumerate}
\item  Scales $r_\theta >1$--15~\hMpc\ are the most relevant scales to constrain
the mean DLA halo mass using the projected cross-correlation $w_{\rm dg}(r_\theta)$.
\item  The DLA-galaxy cross-correlation has an amplitude  $w_{\rm dg}=(\besta)\; w_{\rm gg},$
close to  the predicted value of \biasratiotrue\ using the \citet{MoH_02a} formalism.
\item The inferred mean  DLA halo mass is 
  \begin{equation}
\langle\log M_{\rm DLA}(\msun)\rangle\; = \massDLAlog \;,
 \end{equation} 
 in excellent agreement with
the true values of the simulations, i.e., $\langle \log M_{\rm DLA}\rangle=\massDLAlogtrue$.
Thus, even though DLAs and galaxies
occupy a broad range of halos with massive halos containing multiple galaxies with DLAs,
 the cross-correlation technique yields the first moment of the DLA halo mass distribution.
\item   If we consider subsets of the simulated galaxies with higher star-formation rates
 (representing LBGs), 
the cross-correlation technique is self-consistent, i.e., the DLA mass inferred from the ratio of the correlation functions 
 does not depend on the galaxy sample used.
 This demonstrates the reliability of the method.
\item For real $z=3$ LBGs with a correlation length $r_{\rm 0,gg}\simeq 4$~\hMpc\ \citep[ASSP03;][]{AdelbergerK_04a},
  our results imply that the DLA-LBG cross-correlation is expected to have
a correlation length $r_{\rm 0,dg}\;\simeq \rnotLBG~h^{-1}$~Mpc.
\item With small samples (with 10 lines of sight   and 25 galaxies)
 matching the statistics of BL04,   
  the relative amplitude of the cross-correlation  is $a=0.77\pm0.53$,  i.e.,
 with a  signal-to-noise ratio (S/N$\sim 1.3$--$1.5$) comparable to BL04, where they found $a=1.62\pm1.32$.
\end{enumerate}

In short,
the cross-correlation between galaxies and DLAs is a powerful and self-consistent
technique to constrain the mean mass of DLAs, and
we have demonstrated its reliability.
Given the resolution limits of the simulation used here  ($M_h\ga 5.2\times 10^{10}$~\msun),
 our values are strictly upper limits.
These simulation results  suggest that DLAs are expected to be less massive than $z=3$ LBGs
by a factor of at least $\sim \massratio$.

Recently, \citet{CassataP_04a} studied the morphology of $K$-selected galaxies at redshifts up to $z=2.5$
and found that   the late type fraction drops beyond $z>2$. 
\citet{ErbD_04a}   show that the 
kinematics of 13 $z>2$ morphologically elongated galaxies are not consistent with those of an inclined disk.
Furthermore, the virial mass of these galaxies  is in the range of a few times $10^{10}$~\msun\  up to $5\times 10^{10}$~\msun.
 These results and the ones presented here   disfavor the presence of large, massive $10^{12}~\msun$ disks at $z>2$ and
 therefore the massive disk hypothesis for DLAs.

While current observational samples are just starting to put constraints on $w_{\rm dg}/w_{\rm gg}$
for $z=3$ DLAs---BL04 found $w_{\rm dg}/w_{\rm gg}>\;0$ at the 95\%\ confidence level,
 and ASSP03 found  $<\;1$ at the 90\%\ confidence level, 
allowing the mass range  $\langle\log M_{\rm DLA}\rangle\sim {10}$--${12}$~\msun---
future observations will be able to distinguish between models in which DLAs
 reside in low mass halos from those in which DLAs 
 are massive disks occupying only high mass halos 
 thanks to planned wide-field imagers.

\section*{Acknowledgments}
We thank the anonymous referee for his/her detailed review of the manuscript.
We thank  H. Mo, A. Maller, D. Kere\v{s}, E. Ryan-Weber, and M. Zwaan for many helpful discussions.
  This work was partly supported by the European Community Research and Training Network `The
Physics of the Intergalactic Medium'.  
 J. D. L.  acknowledges support from NSF grant AST 02-06016.

\appendix
\section{A. Cross-correlation and autocorrelation functions} 
\label{appendix:cross}

For a given absorber  with galaxies distributed with ${\mathrm d N}/{\mathrm d z}$,
one may think that  the projected autocorrelation $w_{\rm p,gg}(r_\theta)$
 is proportional to $\int \left(\dNdz\right)^2 \mathrm d z$
while the cross-correlation $w_{\rm dg}(r_\theta)$ is proportional to $\int \left(\dNdz\right)^{1}
\mathrm d z$. Thus, at first glance,  their ratio is  
not very useful. Below we show
the situation to be not so trivial. In this appendix, we merely
connect results previously published to show that the amplitude of 
both $w_{\rm gg}(r_\theta)$ and $w_{\rm dg}(r_\theta)$ are proportional to $1/W_z$, where $W_z$ is the 
redshift width of the galaxy distribution (determined by the box size or by
observational selections such as photometric techniques).
 
First, we list some definitions and three results (eqs.~\ref{croft}--\ref{budavari})
that will be useful later. For a 3D
correlation function $\xi(r)=(r/r_{0})^{-\gamma}$, the projected
correlation function $w_{\rm p}(r_{\rm p})$ is \citep{DavisM_83a}
\begin{eqnarray}
 w_{\rm p}(r_{\rm p})&=&\int_\infty^\infty \mathrm d y \; \xi(r_{\rm p},y)
 		=\int_\infty^\infty \mathrm d y \; \xi(\sqrt{r_{\rm
 		p}^2+y^2}) \,,\nn \\ 
		&=& (r_{\rm p})^{1-\gamma}\; r_{0}^\gamma
 		\; H_\gamma \label{croft} \,,
\end{eqnarray}
where $\xi(r_{\rm p},y)$ is the 3D correlation function decomposed along
the line of sight $y$ and on the plane of the sky $r_{\rm p}$, i.e., $r^2 =
y^2 + r_{\rm p}^2$. The parameter $H_\gamma$ is in fact the beta function
$B(a,b)=\int_0^1 t^{a-1}\;(1-t)^{b-1}\;\mathrm d t$ evaluated with $a=1/2$
and $b=(\gamma-1)/2$,
i.e., $H_\gamma=B({1}/{2}, [{\gamma-1}]/{2})=\Hgamma$.

In appendix C of ASSP03, one finds the expected number of
neighbors between $r_\theta-\mathrm d r/2$ and $r_\theta + \mathrm d r /2$
within a redshift distance $|\Delta_z|\!<\!r_{\rm z}$ 
\begin{eqnarray}
w_{\rm p}(r_{\theta};<\!r_{\rm z})&=&\frac{1}{r_{\rm z}}\int_0^{r_z}
 		\mathrm d l \; \xi(\sqrt{r_{\theta}^2+l^2}) \,,\nn \\
		 &=&
 		\frac{1}{2 r_z} (r_{\theta})^{1-\gamma}\; r_{0}^\gamma \;
 		H_\gamma\; I_x(\frac{1}{2},\frac{\gamma-1}{2})\,,
		\label{adelberger}
\end{eqnarray}
where $x=r_{\rm z}^2/(r_{\rm z}^2+r_\theta^2)$ and $I_x$ is the
incomplete beta function $B_x(a,b)=\int_0^x t^{a-1}\;(1-t)^{b-1}\;\mathrm d
t$ normalized by ${B(a,b)}$: $I_x(a,b)\equiv B_x(a,b)/B(a,b)$.

Many papers \citep{PhillippsS_78a,PeeblesP_93a,BudavariT_03a} have shown
that the angular correlation function is 
\begin{eqnarray}
w(\theta)=(\theta)^{1-\gamma}\; r_{0}^\gamma \; H_\gamma \times
\int_0^\infty \mathrm d z \left(\frac{\mathrm d N}{\mathrm d z}\right)^2 g(z)^{-1} f(z)^{1-\gamma} \,,
\label{budavari}
\end{eqnarray}
where $g(z)=\mathrm d r/\mathrm d z=c/H(z)$ and $f(z)=D_c(z)$ is the
comoving line-of-sight distance to redshift $z$, i.e., $D_c(z)=\int_0^z
\mathrm d t ({c}/{H(t)})$.

Equation~\ref{budavari} can be derived from the definitions of the angular
and 3D correlation functions, $w(\theta)$ and $\xi(r)$
\citep[e.g.,][]{PhillippsS_78a}. We reproduce the derivation here and extend
it to projected auto- and cross-correlation functions. The probabilities of
finding a galaxy in a volume d$V_1$ and another in a volume d$V_2$ at a
distance $ r=|\mathbf r_2\!-\!\mathbf r_1|$  along two lines of sight separated by $\theta$ are 
\begin{eqnarray}
\mathrm d P(\theta)&=&{\cal N}^2\; \mathrm d \Omega_1 \mathrm d \Omega_2
[1+w(\theta)] \label{definition:angular}\,, \\
\end{eqnarray}
or
\begin{eqnarray}
\mathrm d P(r) &=& n(z)^2 \;\mathrm d V_1 \mathrm d V_2 [1+\xi(r)]\,,
\label{definition:xi}
\end{eqnarray}
where $\cal N$ is the number of galaxies per solid angle, i.e., $\mathrm d
N/\mathrm d \Omega$, and $n(z)$ is the number density of galaxies, which
can be a function of redshift. Given that ${\cal N}={1}/{\mathrm d
\Omega} \int \, n(z) \mathrm d V(z)$ and that $\mathrm d V= f^2(z) g(z)
\mathrm \; d \Omega \mathrm d z$, ${\cal N}\equiv \int \mathrm d z\; ({\mathrm d N}/{\mathrm d z}) =
\int \mathrm d z \; n(z) f^2(z)g(z)$.

To relate $w(\theta)$ and $\xi(r)$, one needs to integrate equation~\ref{definition:xi} 
over all possible lines-of-sight separated by $\theta$
(i.e., along $z_1$ and $z_2$) and equate it with
equation~\ref{definition:angular}
\begin{eqnarray}
{\cal N}^2[1+w(\theta)]&=&\int_0^\infty\mathrm d z_1 f(z_1)^2g(z_1)n(z_1)\cdot
\nn \\ && \int_0^\infty d z_2 f(z_2)^2g(z_2)n(z_2)[1+\xi(r_{12})] \;
\,, \label{eq1}
\end{eqnarray}
In the regime of small angles, the distance $r_{12}$ (in comoving Mpc) can
be approximated by
\begin{eqnarray}
r_{12}^2=r_1^2+r_2^2-2r_1r_2 \cos \theta &\simeq&(r_1-r_2)^2+r^2
	\theta^2 \quad \hbox{with}\; r=\frac{r_1+r_2}{2} \;,\nn \\
	&\simeq&(g(z)(z_1-z_2))^2+f(z)^2 \theta^2 \quad \hbox{with}
	z=\frac{z_1+z_2}{2}  \;, \nn\\ 
	&\simeq&g(z)^2y^2+f(z)^2 \theta^2 \quad
	\hbox{with}\; y={z_1-z_2}\, . \label{smallangle}
\end{eqnarray}
Changing variables in equation~\ref{eq1} from ($z_1,z_2$) to ($z,y$),
assuming the the major contribution is from $z_1\simeq z_2$ and using
equation~\ref{smallangle}, the angular correlation function is
\begin{eqnarray}
w(\theta)=\frac{ \int_0^\infty d z f(z)^4g(z)^2n(z)^2\int_{-\infty}^\infty
\mathrm d y \xi(\sqrt{f(z)^2\theta^2+g(z)^2y^2})}{\left[ \int_0^\infty d z
f^2(z)g(z)n(z)\right ]^2}\,. \label{eq:angular}
\end{eqnarray}
Changing variables to $l=g(z)y$, using equation~\ref{croft} and using a
normalized redshift distribution, i.e., $\int \mathrm d z \; ({\mathrm d N}/{\mathrm d z})=1$,
equation~\ref{eq:angular} becomes
\begin{eqnarray}
w(\theta)= \int_0^\infty d z \left( \frac{\mathrm d N}{\mathrm d z} \right)^2 g(z)^{-1} \times [f(z)
\theta]^{1-\gamma}\; r_{0}^\gamma \; H_\gamma \;,\label{intermadiate}
\end{eqnarray}
which leads to equation~\ref{budavari} \citep[eq.~9
in][]{BudavariT_03a} and is one version of Limber's equations.

In this paper, we measured the projected autocorrelation of the LRGs,
$w_{\rm gg}(r_\theta)$, where $r_\theta=f(z)\theta$.\footnote{In general
this should be $D_A (1+z) \theta$ where $D_A$ is the angular distance.  For
a flat universe, $D_A (1+z)=D_M=Dc=f(z)$ where $D_M$ is the comoving
transverse distance, using D. Hogg's notations \citep{HoggD_99a}.}
Following the same steps as above with $r_\theta$ instead of $\theta$, and
$\mathrm d V=(\mathrm d r_\theta)^2 g(z) \mathrm d z$, $w_{\rm
gg}(r_\theta)$ is
\begin{eqnarray}
w_{\rm gg}(r_\theta)= r_\theta^{1-\gamma}\; r_{\rm 0,gg}^\gamma \; H_\gamma
\int_0^\infty d z \left(\frac{\mathrm d N}{\mathrm d z}\right)^2 g(z)^{-1} \;.
\label{auto:result}
\end{eqnarray}
 
In the case of the projected cross-correlation, $w_{\rm dg}(r_\theta)$, the
conditional probability of finding a galaxy in the volume $\mathrm d V_2$
given that there is an absorber at a known position $\mathbf r_1$ is, by
definition \citep[e.g.,][]{EisensteinD_03a},
\begin{eqnarray}
\mathrm d P(2|1)(r_\theta)&=&{\cal N}_{\rm g}\;  \mathrm d \Omega_2 [1+w_{\rm dg}(r_\theta)] \,,  \label{definition2:angular} \\
\mathrm d P(2|1)(r) &=& n_{\rm g}(z) \; \mathrm d V_2 [1+\xi_{\rm dg}(r)] \label{definition2:xi}\,.
\end{eqnarray}
Using the same approximations (eq.~\ref{smallangle}) and one integral
along the line of sight $z_2$  (keeping the absorber at $z_1$), one finds
that the projected cross-correlation is:
\begin{eqnarray}
w_{\rm dg}(r_\theta)&=& \int_0^\infty d z_2 f(z_2)^2g(z_2)n(z_2) \xi_{\rm dg}(r_{12}) \,,\nn \\
&=& \int_0^\infty d z \left(\frac{\mathrm d N}{\mathrm d z} \right)  \xi_{\rm dg}\left(\sqrt{r_\theta^2+g(z)^2(z_1-z_2)^2}\right) \,,\nn\\
&=& \int_0^\infty d y \,g(z) \left(\frac{\mathrm d N}{\mathrm d z} \right) g(z)^{-1} \xi_{\rm dg}\left(\sqrt{r_\theta^2+g(z)^2y^2}\right)\,,\nn\\
&=& \int_0^\infty d l  \frac{\mathrm d N}{\mathrm d l} \xi_{\rm dg}(\sqrt{r_\theta^2+l^2})\label{cross:result}\,,\\
&\simeq&\frac{1}{W_z}\times (r_\theta)^{1-\gamma}\; r_{\rm 0,dg}^\gamma \; H_\gamma \,,
\end{eqnarray}
where we approximated ${\mathrm d N}/{\mathrm d z}$ with a normalized top-hat of width $W_z=\;2\;r_z$,
used  equation~\ref{adelberger}, and the fact that $I_x\simeq 1$, since
$x=r_{\rm z}^2/(r_{\rm z}^2+r_\theta^2)\simeq 1$ for a redshift width $W_z$ of 20~\hMpc\ and $r_\theta=1$~\hMpc.\footnote{The
incomplete beta function $I_x=0.94$ for $W_z=22.222$~\hMpc\ and $r_\theta=1$~\hMpc.}
Thus, as one would have expected, the cross-correlation is inversely
proportional to the width of the galaxy distribution. 

  Naturally, in equations~\ref{definition2:angular} and \ref{definition2:xi}, the redshift
 of galaxy 1 (i.e., the absorber) is assumed to be known with good
 precision. If the absorber population had poorly known redshifts, one would
 need to add an integral to equation~\ref{cross:result}, washing out the
 cross-correlation signal further. 

For the projected autocorrelation (eq.~\ref{auto:result}), if one
approximates $\dNdz$ by a top-hat function of width $W_z$, then
\begin{eqnarray}
w_{\rm gg}(r_\theta)
&=& (r_\theta)^{1-\gamma}\; r_{\rm 0,gg}^\gamma \; H_\gamma  \times \int_0^\infty \mathrm d z \, g(z) \left(\frac{\mathrm d N}{\mathrm d z}\right)^2 g(z)^{-2}
 \,, \nn \\
&=& (r_\theta)^{1-\gamma}\; r_{\rm 0,gg}^\gamma \; H_\gamma  \times \int_0^\infty \mathrm d l \left(\frac{\mathrm d N}{\mathrm d l}\right)^2   
\,,  \nn \\
&\simeq& \left( \frac{1}{W_z}\right )^2 \; W_z \times (r_\theta)^{1-\gamma}\; r_{\rm 0,gg}^\gamma \; H_\gamma \,,
\end{eqnarray}
which shows that the autocorrelation depends on the redshift distribution
of the galaxies in the same way as the cross-correlation, i.e., $\propto
1/W_z$. The reason for this is
 that the redshift distribution ${\mathrm d N}/{\mathrm d z}$ has a very different role with respect to the correlation functions,
 which can be seen by comparing equation~\ref{auto:result} and \ref{cross:result}.
It is this   very different role that leads to the same  $1/W_z$ dependence.

In the case of a Gaussian redshift distribution ${\mathrm d N}/{\mathrm d z}$, the ratio of cross-
and autocorrelations may not be exactly $r_{0, \rm dg}/r_{0,\rm gg}$ if the approximation
leading to A14 breaks down.  Using mock galaxy samples
\citep[from the GIF2 collaboration,][]{GaoL_04a} selected in a redshift
slice of width, $W_z$, equal to their artificial Gaussian redshift errors
$\sigma_z$, we find that the cross-correlation is overestimated by
\overestimate.  This correction factor is
independent of the width of the redshift distribution as long as
$\sigma_z\simeq W_z$  or as long as it is Gaussian.  This implies that the
ratio of the correlation functions ($w_{\rm dg}/w_{\rm gg}$) will be
insensitive to errors in photometric redshifts.

\section{B. The errors to correlation functions}
\label{appendix:errors}

In this appendix, we list the basic properties of the errors to correlation functions.

From the definition of  the cross-correlation $\xi_{dg}$ shown in equation~\ref{eq:cross},
the expected  number of galaxies in a cell of volume $\Delta V$ centered on a DLA
 is given by the counts of neighbor galaxies:
\begin{equation}
  \langle N_{\rm obs}\rangle= \overline N (1+\overline \xi_{dg}(r)),\label{cluster:eq:expected:cross}
\end{equation}
where $\overline N=n_u\Delta V$.

Various text books \citep[e.g.,][section 36]{PeeblesP_80a} have shown that
the   variance of the number of neighbor galaxies $N_{\rm obs}$ near a DLA is the sum of the shot noise, 
\begin{eqnarray}
V(N_{\rm obs})_{\rm sn} &=& N_{\rm obs} \;, \label{cluster:eq:crossvariance:sn}
\end{eqnarray}
and the clustering variance $V(N_{\rm obs})_{\rm cl}$. The clustering variance
is itself the sum of the two terms,  $V_{\rm 2pt}$ and $V_{\rm 3pt}$,
\begin{eqnarray}
V(N_{\rm obs})_{\rm 2pt}&=&\overline N^2 \frac{1}{(\Delta V)^2} \int_{\Delta V} \int_{\Delta V}
    \xi_{\rm gg}(|r_2-r_1|) \label{cluster:eq:crossvariance:2pt}\,,\\
V(N_{\rm obs})_{\rm 3pt}&=&\overline N^2 \frac{1}{(\Delta V)^2} \int_{\Delta V} \int_{\Delta V}
 [\zeta_{\rm dgg}(r_1,r_2)-\xi_{\rm dg}(r_1)\xi_{\rm dg}(r_2)]\mathrm d V_1 \mathrm d V_2  \,,
\label{cluster:eq:crossvariance:3pt}
\end{eqnarray}
where $\overline N=n_u\,\Delta V=N_{\rm exp}$, $\xi_{\rm gg}$ is the  galaxy-galaxy autocorrelation, and 
$\zeta_{\rm dgg}$ is the  three-point correlation function. 
The function $\zeta$ can be written as a product of two-point correlations \citep{PeeblesP_80a},
 \begin{eqnarray}
 \zeta_{\rm dgg}(\mathbf r_1,\mathbf r_2)  &=&Q\;[\xi_{\rm dg}(r_1)\xi_{\rm dg}(r_2)+\xi_{\rm dg}(r_1) \xi_{\rm
 gg}(|\mathbf r_1-\mathbf r_2|)+\xi_{\rm dg}(r_2) \xi_{\rm gg} (|\mathbf r_1-\mathbf  r_2|)], 
  \label{cluster:eq:zeta:anzats}
 \end{eqnarray}
where $r_1=|\mathbf r_o+\mathbf r|$, $r_2=|\mathbf r_o+\mathbf r_2|$ and $r_{12}=|\mathbf r_1-\mathbf r_2|$.

 For a spherical volume $\Delta V$, the integrals
\ref{cluster:eq:crossvariance:2pt} and \ref{cluster:eq:crossvariance:3pt} can be written as 
 \citep[using the results in][\S~59]{PeeblesP_80a}:
\begin{eqnarray}
V(N_{\rm obs})_{\rm 2pt}&=&\overline N^2   \left (\frac{r_{0,\rm gg}}{r}\right )^\gamma\;J_2 
=  \overline N^2 \; \left (\frac{J_2}{K_1}\right) \overline \xi_{\rm gg} \,,\label{cluster:eq:crossvariance:2pt:peebles}\\
V(N_{\rm obs})_{\rm 3pt} &=&\overline N^2  
\left [Q\; \left ( 
K_1^2\left (\frac{r_{0,\rm dg}}{r}\right )^{2\gamma}+
2 K_2\left (\frac{r_{0,\rm dg}}{r}\right )^{\gamma} \left (\frac{r_{0,\rm gg}}{r}\right)^{\gamma}
\right ) - K_1^2 \left (\frac{r_{0,\rm dg}}{r}\right )^{2\gamma}
\right ] \,,  \nonumber \\
&=&\overline N^2  
\left [ Q\; \left (\overline \xi_{\rm dg}^2+2 \frac{K_2}{K_1^2}\overline\xi_{\rm gg}\overline\xi_{\rm dg}  \right ) -
 \overline \xi_{\rm dg}^2 \right ] \,,
\label{cluster:eq:crossvariance:3pt:peebles}
\end{eqnarray}
where $\overline \xi=1/\Delta V\int_0^r \xi \mathrm d V=(r_0/r)^\gamma\;K_1$,
$J_2=72/[(3-\gamma)(4-\gamma)(6-\gamma)2^\gamma]$, $K_1=3/(3-\gamma)$, and $K_2$ can only
be computed numerically. For $\gamma=1.6$, $J_2=4.87$, $K_1=2.14$, and $K_2\simeq 4$.

The variance $V(\xi)$ of the estimator of $\xi$ can be computed analytically. From
\citet{LandyS_93a}, it is 
\begin{eqnarray}
V(\xi)&=&V \left ( \left \langle\frac{N_{\rm obs}}{N_{\rm rand}} \right\rangle \right ) \,,\nn \\
&\simeq &\frac{V(\langle N_{\rm obs}\rangle)}{\langle N_{\rm rand}\rangle^2}+\frac{V(\langle N_{\rm rand}\rangle)\langle N_{\rm obs}\rangle^2}{\langle N_{\rm
rand}\rangle^4} \,, \nn \\
&\simeq &\left [ \frac{V(\langle N_{\rm obs}\rangle)}{\langle N_{\rm obs }\rangle^2}+\frac{V(\langle N_{\rm rand}\rangle)}{\langle N_{\rm rand}\rangle^2 } \right ]
(1+\overline \xi_{dg})^2 \,.\label{cluster:eq:var:xi} 
\end{eqnarray}
The shot noise of the random sample in equation~\ref{cluster:eq:var:xi}, $V(\langle N_{\rm rand}\rangle)/\langle N_{\rm rand}\rangle^2$, 
can be   neglected because the random sample of galaxies is intentionally much larger 
than the sample of observed galaxies.
Thus, the rms (1~$\sigma$) of $\xi_{\rm dg}$ is 
 \begin{eqnarray}
\sigma_{\xi}&\simeq &\frac{\sigma_{\langle N_{\rm obs}\rangle}}{\langle N_{\rm obs}\rangle} [1+\overline \xi_{dg}]  \,,
  \label{cluster:eq:xi:variance}
\end{eqnarray}
  where $\sigma_{\langle N_{\rm obs}\rangle}\equiv \frac{1}{N_{\rm DLA}}\sqrt{ V(N_{\rm obs}) }$, and $V(N_{\rm obs})$
 is given by the sum of equations~\ref{cluster:eq:crossvariance:sn}-\ref{cluster:eq:crossvariance:3pt}.
   If we  approximate the clustering variance of   $N_{\rm obs}$
(eqs.~\ref{cluster:eq:crossvariance:2pt}--\ref{cluster:eq:crossvariance:3pt})
by  $V_{\rm cl}=\overline N^2 (A \overline \xi_{\rm gg}+B \overline \xi^2_{\rm dg}+C\overline \xi_{\rm dg}\overline \xi_{\rm gg} )$,
where  $A=J_2/K_1$, $B= Q-1 $, and $C=2Q K_2/K_1^2$ are  constants
(eqs.~\ref{cluster:eq:crossvariance:2pt:peebles}--\ref{cluster:eq:crossvariance:3pt:peebles}),
then   $\sigma_{\langle N_{\rm obs}\rangle}$  becomes
\begin{eqnarray}
\sigma_{\langle N_{\rm obs}\rangle} &=& \frac{1}{\sqrt{N_{\rm DLA}}}\sqrt{V(N_{\rm obs})_{\rm sn}+V(N_{\rm obs})_{\rm cl}} \,,\nn \\
	&=&  \frac{1}{\sqrt{N_{\rm DLA}}}\sqrt{\langle N_{\rm obs}\rangle+\overline N^2 
	( A \overline \xi_{\rm gg}+B \overline \xi^2_{\rm dg}+C\overline \xi_{\rm dg}\overline \xi_{\rm gg})
	} , \label{cluster:eq:Nobs:variance}
\end{eqnarray}
 Therefore, 
the expected rms of the cross-correlation function  $\sigma_{w}$, equation~\ref{cluster:eq:xi:variance}, becomes 
\begin{eqnarray}
\sigma_{\xi}&=&\frac{1}{\sqrt {N_{\rm DLA}}}\frac{1}{\sqrt{\langle N_{\rm obs}\rangle}} 
\; \sqrt{1+\overline N^2\frac{A\overline \xi_{gg}+B \overline \xi^2_{dg}+C\overline \xi_{\rm dg}\overline \xi_{\rm gg}}{\langle N_{\rm 	obs}\rangle} } 	
\;  (1+\overline \xi_{dg}) \,,  \nonumber
\end{eqnarray}
or 
\begin{eqnarray}
\sigma_{\xi}&=&\frac{1}{\sqrt {N_{\rm DLA}}}\frac{1}{\sqrt{\overline N}} \;  
\sqrt{1+ \frac{A\overline \xi_{gg}+B \overline \xi^2_{dg}+C\overline \xi_{\rm dg}\overline \xi_{\rm gg}}{  (1+\overline \xi_{dg})} } 	 
  \; \sqrt{ (1+\overline \xi_{dg})} \,, \label{cluster:eq:crossxi:variance} 
\end{eqnarray}
using equation~\ref{cluster:eq:expected:cross}.
This expression is proportional to $\frac{1}{\sqrt {N_{\rm DLA}}}\frac{1}{\sqrt{N}}$
  as one might have expected.
Thus, the noise in  $\langle\xi\rangle$ goes as the inverse of the square
root of the number of DLAs, $N_{\rm DLA}$, and as the inverse of the square root of
the number of galaxies  $N$ in the cell of volume $\Delta V$.



\begin{thebibliography}{70}
\small
\itemindent -0.48cm
\expandafter\ifx\csname natexlab\endcsname\relax\def\natexlab#1{#1}\fi

\bibitem[{{Adelberger} {et~al.}(2004){Adelberger}, {Steidel}, {Pettini},
  {Shapley}, {Reddy}, \& {Erb}}]{AdelbergerK_04a}
{Adelberger}, K.~L., {Steidel}, C.~C., {Pettini}, M., {Shapley}, A.~E.,
  {Reddy}, N.~A., \& {Erb}, D.~K. 2004, \apj, accepted, astro-ph/0410165

\bibitem[{{Adelberger} {et~al.}(2003){Adelberger}, {Steidel}, {Shapley}, \&
  {Pettini}}]{AdelbergerK_03a}
{Adelberger}, K.~L., {Steidel}, C.~C., {Shapley}, A.~E., \& {Pettini}, M. 2003,
  \apj, 584, 45

\bibitem[{{Berlind} \& {Weinberg}(2002)}]{BerlindA_02a}
{Berlind}, A.~A., \& {Weinberg}, D.~H. 2002, \apj, 575, 587

\bibitem[{{Berlind} {et~al.}(2003){Berlind}, {Weinberg}, {Benson}, {Baugh},
  {Cole}, {Dav{\' e}}, {Frenk}, {Jenkins}, {Katz}, \& {Lacey}}]{BerlindA_03a}
{Berlind}, A.~A. {et~al.} 2003, \apj, 593, 1

\bibitem[{{Bernstein}(1994)}]{BernsteinG_94a}
{Bernstein}, G.~M. 1994, \apj, 424, 569

\bibitem[{{Bouch{\' e}}(2003)}]{BoucheN_03b}
{Bouch{\' e}}, N. 2003, PhD thesis, Univ. Massachusetts, Amherst

\bibitem[{{Bouch{\' e}} \& {Lowenthal}(2003)}]{BoucheN_03a}
{Bouch{\' e}}, N., \& {Lowenthal}, J.~D. 2003, \apj, 596, 810

\bibitem[{{Bouch{\' e}} \& {Lowenthal}(2004)}]{BoucheN_04c}
---. 2004, \apj, 609, 513

\bibitem[{{Bouch{\' e}} {et~al.}(2004){Bouch{\' e}}, {Murphy}, \& {P{\'
  e}roux}}]{BoucheN_04a}
{Bouch{\' e}}, N., {Murphy}, M.~T., \& {P{\' e}roux}, C. 2004, \mnras, 354, 25L

\bibitem[{{Budav{\' a}ri}~{et~al.}(2003)}]{BudavariT_03a}
{Budav{\' a}ri}, T., \& {et~al.,}. 2003, \apj, 595, 59

\bibitem[{{Cassata} {et~al.}(2004){Cassata}, {Cimatti}, {Franceschini},
  {Daddi}, {Pignatelli}, {Fasano}, {Rodighiero}, {Pozzetti}, {Mignoli}, \&
  {Renzini}}]{CassataP_04a}
{Cassata}, P. {et~al.} 2004, \mnras, accepted (astro-ph/0411768)

\bibitem[{{Cooke} {et~al.}(2005){Cooke}, {Wolfe}, {Prochaska}, \&
  {Gawiser}}]{CookeJ_05a}
{Cooke}, J., {Wolfe}, A.~M., {Prochaska}, J.~X., \& {Gawiser}, E. 2005, \apj, 621,
596

\bibitem[{{Dav{\' e}} {et~al.}(1999){Dav{\' e}}, {Hernquist}, {Katz}, \&
  {Weinberg}}]{DaveR_99a}
{Dav{\' e}}, R., {Hernquist}, L., {Katz}, N., \& {Weinberg}, D.~H. 1999, \apj,
  511, 521

\bibitem[{{Dav{\'e}} {et~al.}(1997){Dav{\'e}}, {Dubinski}, \&
  {Hernquist}}]{DaveR_97b}
{Dav{\'e}}, R., {Dubinski}, J., \& {Hernquist}, L. 1997, New Astronomy, 2, 277

\bibitem[{{Davis} {et~al.}(1985){Davis}, {Efstathiou}, {Frenk}, \&
  {White}}]{DavisM_85a}
{Davis}, M., {Efstathiou}, G., {Frenk}, C.~S., \& {White}, S.~D.~M. 1985, \apj,
  292, 371

\bibitem[{{Davis} \& {Peebles}(1983)}]{DavisM_83a}
{Davis}, M., \& {Peebles}, P.~J.~E. 1983, \apj, 267, 465

\bibitem[{{D'Odorico} {et~al.}(2002){D'Odorico}, {Petitjean}, \&
  {Cristiani}}]{DodoricoV_02a}
{D'Odorico}, V., {Petitjean}, P., \& {Cristiani}, S. 2002, \aap, 390, 13

\bibitem[{{Efron}(1982)}]{EfronB_82a}
{Efron}, B. 1982, {The Jackknife, the Bootstrap and other resampling plans}
  (Philadelphia, U.S.A.: Society for Industrial and Applied Mathematics (SIAM))

\bibitem[{{Eisenstein}(2003)}]{EisensteinD_03a}
{Eisenstein}, D.~J. 2003, \apj, 586, 718

\bibitem[{{Ellison} {et~al.}(2001){Ellison}, {Pettini}, {Steidel}, \&
  {Shapley}}]{EllisonS_01a}
{Ellison}, S.~L., {Pettini}, M., {Steidel}, C.~C., \& {Shapley}, A.~E. 2001,
  \apj, 549, 770

\bibitem[{{Erb} {et~al.}(2004){Erb}, {Steidel}, {Shapley}, {Pettini}, \&
  {Adelberger}}]{ErbD_04a}
{Erb}, D.~K., {Steidel}, C.~C., {Shapley}, A.~E., {Pettini}, M., \&
  {Adelberger}, K.~L. 2004, \apj, 612, 122

\bibitem[{{Francis} \& {Hewett}(1993)}]{FrancisP_93a}
{Francis}, P.~J., \& {Hewett}, P.~C. 1993, \aj, 105, 1633

\bibitem[{{Fynbo} {et~al.}(2003){Fynbo}, {Ledoux}, {M{\" o}ller}, {Thomsen}, \&
  {Burud}}]{FynboJ_03a}
{Fynbo}, J.~P.~U., {Ledoux}, C., {M{\" o}ller}, P., {Thomsen}, B., \& {Burud},
  I. 2003, \aap, 407, 147

\bibitem[{{Fynbo} {et~al.}(1999){Fynbo}, {M{\o}ller}, \& {Warren}}]{FynboJ_99a}
{Fynbo}, J.~U., {M{\o}ller}, P., \& {Warren}, S.~J. 1999, \mnras, 305, 849

\bibitem[{{Gao} {et~al.}(2004){Gao}, {White}, {Jenkins}, {Stoehr}, \&
  {Springel}}]{GaoL_04a}
{Gao}, L., {White}, S.~D.~M., {Jenkins}, A., {Stoehr}, F., \& {Springel}, V.
  2004, \mnras, 355, 819

\bibitem[{{Gardner} {et~al.}(1997{\natexlab{a}}){Gardner}, {Katz}, {Hernquist},
  \& {Weinberg}}]{GardnerJ_97a}
{Gardner}, J.~P., {Katz}, N., {Hernquist}, L., \& {Weinberg}, D.~H.
  1997{\natexlab{a}}, \apj, 484, 31

\bibitem[{{Gardner} {et~al.}(2001){Gardner}, {Katz}, {Hernquist}, \&
  {Weinberg}}]{GardnerJ_01a}
---. 2001, \apj, 559, 131

\bibitem[{{Gardner} {et~al.}(1997{\natexlab{b}}){Gardner}, {Katz}, {Weinberg},
  \& {Hernquist}}]{GardnerJ_97b}
{Gardner}, J.~P., {Katz}, N., {Weinberg}, D.~H., \& {Hernquist}, L.
  1997{\natexlab{b}}, \apj, 486, 42

\bibitem[{{Gawiser} {et~al.}(2001){Gawiser}, {Wolfe}, {Prochaska}, {Lanzetta},
  {Yahata}, \& {Quirrenbach}}]{GawiserE_01a}
{Gawiser}, E., {Wolfe}, A.~M., {Prochaska}, J.~X., {Lanzetta}, K.~M., {Yahata},
  N., \& {Quirrenbach}, A. 2001, \apj, 562, 628

\bibitem[{{Haardt} \& {Madau}(1996)}]{HaardtF_96a}
{Haardt}, F., \& {Madau}, P. 1996, \apj, 461, 20

\bibitem[{{Haehnelt} {et~al.}(1998){Haehnelt}, {Steinmetz}, \&
  {Rauch}}]{HaehneltM_98a}
{Haehnelt}, M.~G., {Steinmetz}, M., \& {Rauch}, M. 1998, \apj, 495, 647

\bibitem[{{Haehnelt} {et~al.}(2000){Haehnelt}, {Steinmetz}, \&
  {Rauch}}]{HaehneltM_00a}
---. 2000, \apj, 534, 594

\bibitem[{{Hernquist}(1987)}]{HernquistL_87a}
{Hernquist}, L. 1987, \apjs, 64, 715

\bibitem[{Hogg(1999)}]{HoggD_99a}
Hogg, D.~W. 1999, preprint (astro-ph/9905116)

\bibitem[{{Katz} {et~al.}(1996{\natexlab{a}}){Katz}, {Weinberg}, \&
  {Hernquist}}]{KatzN_96b}
{Katz}, N., {Weinberg}, D.~H., \& {Hernquist}, L. 1996{\natexlab{a}}, \apjs,
  105, 19

\bibitem[{{Katz} {et~al.}(1996{\natexlab{b}}){Katz}, {Weinberg}, {Hernquist},
  \& {Miralda-Escude}}]{KatzN_96a}
{Katz}, N., {Weinberg}, D.~H., {Hernquist}, L., \& {Miralda-Escude}, J.
  1996{\natexlab{b}}, \apjl, 457, L57

\bibitem[{{Kauffmann}(1996)}]{KauffmannG_96a}
{Kauffmann}, G. 1996, \mnras, 281, 475

\bibitem[{{Kere\v{s}} {et~al.}(2004){Kere\v{s}}, {Katz}, {Weinberg}, \&
  {Dav{\'e}}}]{KeresD_04a}
{Kere\v{s}}, D., {Katz}, N., {Weinberg}, D.~H., \& {Dav{\'e}}, R. 2004, \mnras,
  submitted, preprint (astro-ph/0407095)

\bibitem[{{Kulkarni} {et~al.}(2000){Kulkarni}, {Hill}, {Schneider}, {Weymann},
  {Storrie-Lombardi}, {Rieke}, {Thompson}, \& {Jannuzi}}]{KulkarniV_00a}
{Kulkarni}, V.~P., {Hill}, J.~M., {Schneider}, G., {Weymann}, R.~J.,
  {Storrie-Lombardi}, L.~J., {Rieke}, M.~J., {Thompson}, R.~I., \& {Jannuzi},
  B.~T. 2000, \apj, 536, 36

\bibitem[{{Landy} \& {Szalay}(1993)}]{LandyS_93a}
{Landy}, S.~D., \& {Szalay}, A.~S. 1993, \apj, 412, 64

\bibitem[{{Lanzetta} {et~al.}(1991){Lanzetta}, {McMahon}, {Wolfe}, {Turnshek},
  {Hazard}, \& {Lu}}]{LanzettaK_91a}
{Lanzetta}, K.~M., {McMahon}, R.~G., {Wolfe}, A.~M., {Turnshek}, D.~A.,
  {Hazard}, C., \& {Lu}, L. 1991, \apjs, 77, 1

\bibitem[{{Lanzetta} {et~al.}(1995){Lanzetta}, {Wolfe}, \&
  {Turnshek}}]{LanzettaK_95b}
{Lanzetta}, K.~M., {Wolfe}, A.~M., \& {Turnshek}, D.~A. 1995, \apj, 440, 435

\bibitem[{{Le Brun} {et~al.}(1997){Le Brun}, {Bergeron}, {Boisse}, \&
  {Deharveng}}]{LeBrunV_97a}
{Le Brun}, V., {Bergeron}, J., {Boisse}, P., \& {Deharveng}, J.~M. 1997, \aap,
  321, 733

\bibitem[{{Ledoux} {et~al.}(1998){Ledoux}, {Petitjean}, {Bergeron}, {Wampler},
  \& {Srianand}}]{LedouxC_98a}
{Ledoux}, C., {Petitjean}, P., {Bergeron}, J., {Wampler}, E.~J., \& {Srianand},
  R. 1998, \aap, 337, 51

\bibitem[{{Lowenthal} {et~al.}(1991){Lowenthal}, {Hogan}, {Green}, {Caulet},
  {Woodgate}, {Brown}, \& {Foltz}}]{LowenthalJ_91a}
{Lowenthal}, J.~D., {Hogan}, C.~J., {Green}, R.~F., {Caulet}, A., {Woodgate},
  B.~E., {Brown}, L., \& {Foltz}, C.~B. 1991, \apjl, 377, L73

\bibitem[{{Lowenthal} {et~al.}(1995){Lowenthal}, {Hogan}, {Green}, {Woodgate},
  {Caulet}, {Brown}, \& {Bechtold}}]{LowenthalJ_95a}
{Lowenthal}, J.~D., {Hogan}, C.~J., {Green}, R.~F., {Woodgate}, B., {Caulet},
  A., {Brown}, L., \& {Bechtold}, J. 1995, \apj, 451, 484

\bibitem[{{Lowenthal} {et~al.}(1997){Lowenthal}, {Koo}, {Guzman}, {Gallego},
  {Phillips}, {Faber}, {Vogt}, {Illingworth}, \& {Gronwall}}]{LowenthalJ_97a}
{Lowenthal}, J.~D. {et~al.} 1997, \apj, 481, 673

\bibitem[{{Lucy}(1977)}]{LucyL_77a}
{Lucy}, L.~B. 1977, \aj, 82, 1013

\bibitem[{{Maller} {et~al.}(2000){Maller}, {Prochaska}, {Somerville}, \&
  {Primack}}]{MallerA_00a}
{Maller}, A., {Prochaska}, J., {Somerville}, R., \& {Primack}, J. 2000, in ASP
  Conf. Ser. 200: Clustering at High Redshift, 430

\bibitem[{{McDonald} \& {Miralda-Escud{\' e}}(1999)}]{McDonaldP_99a}
{McDonald}, P., \& {Miralda-Escud{\' e}}, J. 1999, \apj, 519, 486

\bibitem[{{Miller} \& {Scalo}(1979)}]{MillerG_79a}
{Miller}, G.~E., \& {Scalo}, J.~M. 1979, \apjs, 41, 513

\bibitem[{{Mo} {et~al.}(1999){Mo}, {Mao}, \& {White}}]{MoH_99a}
{Mo}, H.~J., {Mao}, S., \& {White}, S.~D.~M. 1999, \mnras, 304, 175

\bibitem[{{Mo} {et~al.}(1993){Mo}, {Peacock}, \& {Xia}}]{MoH_93a}
{Mo}, H.~J., {Peacock}, J.~A., \& {Xia}, X.~Y. 1993, \mnras, 260, 121

\bibitem[{{Mo} \& {White}(2002)}]{MoH_02a}
{Mo}, H.~J., \& {White}, S.~D.~M. 2002, \mnras, 336, 112

\bibitem[{{Nagamine} {et~al.}(2004){Nagamine}, {Springel}, \&
  {Hernquist}}]{NagamineK_04a}
{Nagamine}, K., {Springel}, V., \& {Hernquist}, L. 2004, \mnras, 348, 421

\bibitem[{{Okoshi} {et~al.}(2004){Okoshi}, {Nagashima}, {Gouda}, \&
  {Yoshioka}}]{OkoshiK_04a}
{Okoshi}, K., {Nagashima}, M., {Gouda}, N., \& {Yoshioka}, S. 2004, \apj, 603,
  12

\bibitem[{{P{\' e}roux} {et~al.}(2003){P{\' e}roux}, {Dessauges-Zavadsky},
  {D'Odorico}, {Kim}, \& {McMahon}}]{PerouxC_03b}
{P{\' e}roux}, C., {Dessauges-Zavadsky}, M., {D'Odorico}, S., {Kim}, T., \&
  {McMahon}, R.~G. 2003, \mnras, 345, 480

\bibitem[{{Peebles}(1980)}]{PeeblesP_80a}
{Peebles}, P.~J.~E. 1980, {The large-scale structure of the universe} (Research
  supported by the National Science Foundation.~Princeton, N.J., Princeton
  University Press, 1980.~435 p.)

\bibitem[{{Peebles}(1993)}]{PeeblesP_93a}
---. 1993, {Principles of physical cosmology} (Princeton, NJ, USA: Princeton
  University Press)

\bibitem[{{Pettini} {et~al.}(2001){Pettini}, {Shapley}, {Steidel}, {Cuby},
  {Dickinson}, {Moorwood}, {Adelberger}, \& {Giavalisco}}]{PettiniM_01a}
{Pettini}, M., {Shapley}, A.~E., {Steidel}, C.~C., {Cuby}, J., {Dickinson}, M.,
  {Moorwood}, A.~F.~M., {Adelberger}, K.~L., \& {Giavalisco}, M. 2001, \apj,
  554, 981

\bibitem[{{Phillipps} {et~al.}(1978){Phillipps}, {Fong}, {Fall}, \&
  {MacGillivray}}]{PhillippsS_78a}
{Phillipps}, S., {Fong}, R., {Fall}, R.~S.~E.~S.~M., \& {MacGillivray}, H.~T.
  1978, \mnras, 182, 673

\bibitem[{{Prochaska} \& {Wolfe}(1997{\natexlab{a}})}]{ProchaskaJ_97a}
{Prochaska}, J.~X., \& {Wolfe}, A.~M. 1997{\natexlab{a}}, \apj, 474, 140

\bibitem[{{Prochaska} \& {Wolfe}(1997{\natexlab{b}})}]{ProchaskaJ_97b}
---. 1997{\natexlab{b}}, \apj, 487, 73

\bibitem[{{Rao} \& {Turnshek}(2000)}]{RaoS_00a}
{Rao}, S.~M., \& {Turnshek}, D.~A. 2000, \apjs, 130, 1

\bibitem[{{Roche} {et~al.}(2000){Roche}, {Lowenthal}, \&
  {Woodgate}}]{RocheN_00a}
{Roche}, N., {Lowenthal}, J., \& {Woodgate}, B. 2000, \mnras, 317, 937

\bibitem[{{Rosenberg} \& {Schneider}(2003)}]{RosenbergJ_03a}
{Rosenberg}, J.~L., \& {Schneider}, S.~E. 2003, \apj, 585, 256

\bibitem[{{Schaye}(2001)}]{SchayeJ_01a}
{Schaye}, J. 2001, \apjl, 559, L1

\bibitem[{{Wolfe}(1993)}]{WolfeA_93a}
{Wolfe}, A.~M. 1993, \apj, 402, 411

\bibitem[{{Wolfe} {et~al.}(1995){Wolfe}, {Lanzetta}, {Foltz}, \&
  {Chaffee}}]{WolfeA_95a}
{Wolfe}, A.~M., {Lanzetta}, K.~M., {Foltz}, C.~B., \& {Chaffee}, F.~H. 1995,
  \apj, 454, 698

\bibitem[{{Wolfe} {et~al.}(1986){Wolfe}, {Turnshek}, {Smith}, \&
  {Cohen}}]{WolfeA_86a}
{Wolfe}, A.~M., {Turnshek}, D.~A., {Smith}, H.~E., \& {Cohen}, R.~D. 1986,
  \apjs, 61, 249

\bibitem[{{Zwaan} {et~al.}(2005){Zwaan}, {van der Hulst}, {Briggs}, {Verheijen}, \& {Ryan-Weber}}]{ZwaanM_05a}
{Zwaan}, M.~A., {van der Hulst}, J.~M., {Briggs}, F.~H., {Verheijen}, M.~A.~W., \& {Ryan-Weber}, E.~V. 2005, \mnras, submitted

\end{thebibliography}


\end{document}